\documentclass[reprint,amsmath,amssymb,pra,aps,groupedaddress,showpacs,nofootinbib]{revtex4-1}
\usepackage{bm,dcolumn,graphicx,epsfig,color,comment,epstopdf}
\usepackage{ulem,amssymb,amsmath,subeqnarray,mathrsfs,
           amsfonts,bm,makecell,cases,float,extpfeil,lipsum}
\usepackage[colorlinks,linkcolor=blue,anchorcolor=blue,citecolor=blue,
bookmarksnumbered
]{hyperref}
\allowdisplaybreaks[4]
\def\a{\alpha}
\def\b{\beta}

\def\d{\delta}
\def\D{\Delta}
\def\s{\sigma}

\begin{document}
%%%%%%%%%%%%%%%%%%%%%%%%%%%%%%%%%%%%%%%%%%%%%%%%%%%%%%%%%%%%%%%%%%%%%%%%%%%%%%%%%
\preprint{APS/123-QED}

\title{Giant Kerr nonlinearities and magneto-optical rotations in a Rydberg-atom gas via double electromagnetically induced transparency}
\author{Yue Mu$^{1}$, Lu Qin$^{1}$, Zeyun Shi$^{1}$, and Guoxiang Huang$^{1,2,3}$}
%\email{gxhuang@phy.ecnu.edu.cn}
\affiliation{$^1$State Key Laboratory of Precision Spectroscopy, East China Normal University, Shanghai 200062, China\\
$^2$NYU-ECNU Joint Institute of Physics, New York University at Shanghai, Shanghai 200062, China\\
$^3$Collaborative Innovation Center of Extreme Optics, Shanxi University, Taiyuan 030006, China}

\date{\today}

\begin{abstract}
We investigate the Kerr and magneto-optical effects for a probe laser field with two orthogonally polarized components, propagating in a cold Rydberg atomic gas with an inverted-Y-type level configuration via double electromagnetically induced transparency (EIT). Through an approach beyond both mean-field and ground-state approximations, we make detailed calculations on third-order nonlinear optical susceptibilities and show that the system possesses giant nonlocal self- and cross-Kerr nonlinearities contributed by Rydberg-Rydberg interaction. The theoretical result of the cross-Kerr nonlinearity obtained for $^{85}$Rb atomic gas is very close to the experimental one reported recently. Moreover, we demonstrate that the probe laser field can acquire a very large magneto-optical rotation via the double EIT, which may be used to design atomic magnetometers with high precision. The results presented here are promising not only for the development of nonlocal nonlinear magneto-optics but also for applications in precision measurement and optical information processing and transmission based on Rydberg atomic gases.

DOI:\href{https://doi.org/10.1103/PhysRevA.00.003700}{10.1103/PhysRevA.00.003700}
%% 42.50.Gy (EIT), 42.65.An (Optical susceptibility), 32.80.Ee (Rydberg state), 07.55.Ge (Magnetometers for magnetic field measurements)

\end{abstract}

\maketitle

%\tableofcontents

%%%%%%%%%%%%%%%%%%%%%%%%%%%%%%%%%%%%%%%%%%%%%%%%%%%%%%%%%%%%%%%%%%%%%%
\section{Introduction}\label{sec1}

The study of the Kerr effect is a key topic in nonlinear optics because the Kerr effect is essential for the realization of most nonlinear optical processes~\cite{Shen1984,Boyd2008}. In recent decades, tremendous new applications of the Kerr effect have been found such as quantum optical squeezing~\cite{Andersen2016,Schnabel2017}, quantum entanglement and concentration~\cite{Kwiat1995,Fiurasek2003,Shih2003,Pan2013,Tatham2014}, quantum nondemolition measurements~\cite{Imoto1985,Roch1997,Grangier1998}, single-photon switches and transistors~\cite{Chang2014-1}, and quantum computation and quantum information~\cite{Milburn1989,Chuang1995,Turchette1995,Nielsen2000,Vitali2000,
Ottaviani2003,Nemoto2004,Munro2005,Rebic2006-1,Kok2008,Lin2009,Hang2010,Li2013}. Usually, passive optical media (e.g., glass-based optical fibers) are exploited for generating Kerr nonlinearity, where excitation schemes are far-off-resonance ones for evading high optical absorption. The Kerr nonlinearity realized in this way is weak and thus cannot meet the ever-increasing demand for optical information processing and transmission.

To obtain a large Kerr nonlinearity, a natural idea is to make use of active (resonant) optical media, which however results in significant optical absorption. One of the methods to resolve this problem is the utilization of electromagnetically induced transparency (EIT), typically occurring in resonant three-level atomic systems where the optical absorption of a probe laser field can be largely suppressed by the quantum destruction effect induced by a control laser field~\cite{Harris1997}. In addition to the suppression of optical absorption, the light
propagation in EIT media exhibits also many other interesting properties, including the significant reduction of group velocity and the resonant enhancement of Kerr nonlinearity~\cite{Fleischhauer2005,Khurgin2009}, by which important applications (e.g., photonic memory, quantum phase gates, entangled photon sources, optical clocks, highly efficient four-wave mixing, optical splitters and routers,
% highly sensitive magnetometers,
and slow-light solitons) can be realized~\cite{Lvovsky2009,Simon2010,Sangouard2011,Bussieres2013,Heshami2016,
Ottaviani2003,Rebic2006-1,Kok2008,Lin2009,Hang2010,Li2013,
Wal2003,Kuzmich2003,Balic2005,Du2008,Santra2005,Zanon2006,Fleischhauer2005,
Khurgin2009,Wang2004,Raczynski2007,Xiao2008,Yang2015,Shou2019,
Wu2004,Huang2005,Chen2014}. Nevertheless, the Kerr nonlinearity obtained in conventional EIT media is still too small for many nonlinear optical processes working at single-photon levels.

In recent years, considerable attention has been paid to the investigation of cold Rydberg atomic gases, where atoms are electrically excited to quantum states with a very large principal quantum number $n$ (i.e., Rydberg states) which possess many striking features~\cite{Gallagher2008,Saffman2010,Adams2020}. One of the research directions in this vibrant field is the study of nonlinear and quantum optical effects based on Rydberg EIT~\cite{Mohapatra2007,Pritchard2010}, where three levels with a ladder-type configuration are employed, and the Kerr nonlinearity in such systems has been investigated both experimentally and theoretically. It has been shown that the Kerr nonlinearity via the Rydberg EIT can be enhanced several orders of magnitude compared to conventional EIT~\cite{Pritchard2011,Sevincli2011a,Ates2011,Parigi2012,
Stanojevic2013,Grankin2015,Boddeda2016,Bienias2016,Bai2016,
Tebben2019,Bai2019}. The reason is that the contribution to the nonlinear optical susceptibilities by the interaction between Rydberg atoms (called Rydberg-Rydberg interaction) is much larger than cases where the Rydberg-Rydberg interaction is absent~\cite{Bai2016,Bai2019,note000}\footnote{If the Rydberg-Rydberg interaction is absent, EIT systems may support local Kerr nonlinearity with small absorption that occurs by photon-atom interaction when the two-photon detuning $\Delta_4\neq 0$. See \cite{Bai2016,Bai2019,note000}.}.

Recently, Sinclair {\it et al.} reported the first experimental observation of cross-Kerr nonlinearity in a cold $^{85}$Rb atomic gas with an inverted-Y-type level configuration via a double Rydberg EIT~\cite{Sinclair2019}. Through the measurement of a nonlinear phase written onto a probe laser pulse, they found that due to the Rydberg-Rydberg interaction the third-order nonlinear optical susceptibility $\chi^{(3)}$ of the system can reach the order of magnitude $1\times 10^{-8}\,{\rm m}^{2}{\rm V}^{-2}$. Because cross-Kerr nonlinearities have potential applications ranging from optical quantum information processing to quantum nondemolition measurement, it is necessary and timely to make a detailed theoretical study of the self- and cross-Kerr nonlinear effects in systems working with the double Rydberg EIT.

In this work, we investigate theoretically the Kerr nonlinearity in a cold, inverted-Y-type atomic gas working under the condition of double Rydberg EIT. We assume that the two lower levels of the atoms are Zeeman sublevels (split from a hyperfine ground-state level by a weak, static magnetic field) and coupled by a probe laser field with two orthogonally polarized components [see Fig.~\ref{Fig1}(a)\,]. By means of an approach beyond both the mean-field approximation (MFA) and the ground-state approximation (GSA)~\cite{note001}
\footnote{The so-called ground-state approximation (GSA) is the one in which one assumes that the diagonal elements of one-atom density matrix (i.e., atomic populations) $\rho_{\alpha\alpha}$ are assumed to keep the values of their initial preparations during the time evolution of the system (see~\cite{note001}). Such an approximation was widely used in the study of conventional EIT and was also employed by some authors for calculations of the Kerr nonlinearities of Rydberg atomic gases; see, e.g.,  Refs.~\cite{Sevincli2011a,Bienias2016,Tebben2019}.},
we present systematic and detailed calculations of the third-order nonlinear optical susceptibilities of the system. We show that such a system possesses giant nonlocal self-Kerr and cross-Kerr nonlinearities contributed by the Rydberg-Rydberg interaction. Our theoretical result on the cross-Kerr nonlinearity of $^{85}$Rb atomic gas is very close to the experimental measurement reported by Sinclair {\it et al}.~\cite{Sinclair2019}.
Moreover, we demonstrate that, by virtue of the double Rydberg EIT, the probe field may acquire a very large magneto-optical rotation (MOR) if a very weak external magnetic field is applied.

%%%%%%%%%%%%%%%%%%%
Before proceeding, we would like to emphasize that, although in recent years a number of theoretical studies on the Kerr nonlinearity in Rydberg atomic gases have appeared~\cite{Sevincli2011a,Ates2011,Stanojevic2013,Grankin2015,Boddeda2016,Bienias2016,Bai2016,
Tebben2019,Bai2019}, our work calculates the self- and cross-Kerr nonlinearities via double EIT beyond the mean-field approximation. Furthermore, our calculated result agrees well with the experimental observation~\cite{Sinclair2019}, which will possibly trigger further theoretical and experimental investigations because the giant cross-Kerr nonlinearities of Rydberg gases have important applications in optical and quantum information processing. Moreover, the method of calculation developed here has not only given reasonable results agreeing with the experiment, but also clarified the confusion in the literature where several theoretical approaches [i.e. MFA, GSA, and reduced density matrix expansion (RDME); see Secs.~\ref{sec30} and \ref{sec33}]  were adopted for the calculation of Kerr nonlinearities in Rydberg gases. In addition, the giant MOR predicted here is one order of magnitude larger than that obtained by using conventional EIT (see Sec.~\ref{sec4}), which can be used to design atomic magnetometers with much higher precision. Thus, the research presented here opens a route for the development of nonlocal nonlinear magneto-optics and the results are promising for practical applications in precision measurements, optical information processing and transmission based on Rydberg atomic gases~\cite{Pritchard2013-1,Firstenberg2016-1,Murray2016-1}.

The remainder of the paper is arranged as follows. In Sec.~\ref{sec2} the physical model of the double Rydberg EIT is described and its linear optical property is discussed. In Sec.~\ref{sec3} the third-order nonlinear optical susceptibilities of the system in regimes of both dispersion and dissipation are calculated in detail beyond the MFA and GSA.  In Sec.~\ref{sec4} coupled envelope equations for the two polarization components of the probe field are derived and the giant MOR and the possibility of realizing highly sensitive magnetometers are explored. Section \ref{sec5} contains a summary of the main results obtained in this work. Some explicit expressions of equations of motion and related calculation details for finding their solutions are given in Appendixes \ref{appendix0}-\ref{appd}.

%%%%%%%%%%%%%%%%%%%%%%%%%%%%
\section{Model and linear dispersion relation}\label{sec2}

\subsection{Model}

\begin{figure}
\centering
\includegraphics[width=0.98\linewidth]{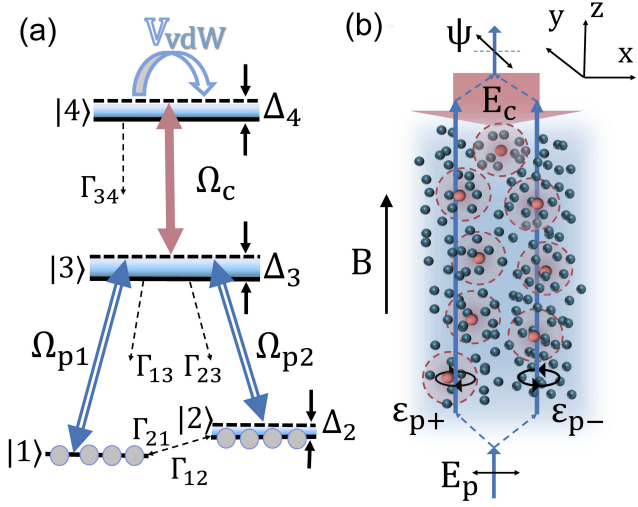}
\caption{\footnotesize  Schematics of the model.
(a)~Level diagram and excitation scheme of the double Rydberg EIT, realized by an inverted-Y-type four-level atomic gas. Here $|1\rangle$ and $|2\rangle$ are two ground states (split from a hyperfine ground-state level by a weak static magnetic field $B$) and $|3\rangle$ and $|4\rangle$ are the intermediate state and the high-lying Rydberg state, respectively. The two circularly polarized components of the probe field (with half-Rabi frequencies $\Omega_{{p}1}$ and $\Omega_{{p}2}$ respectively) couple the transitions
$|1\rangle\leftrightarrow|3\rangle$ and $|2\rangle\leftrightarrow|3\rangle$, respectively. The control field (with the half-Rabi frequency $\Omega_{{c}}$) couples the transition $|3\rangle\leftrightarrow|4\rangle$. In addition,
$\Delta_j$ are detunings; $\Gamma_{jl}$ are rates of spontaneous decay or incoherent population transfer. The interaction between the two Rydberg atoms located at positions ${\bf r}$ and ${\bf r}'$, respectively, is described by the van der Waals potential $V_{\rm{vdW}}\equiv\hbar V({\bf r}^{\prime}-{\bf r})$ with $V({\bf r}^{\prime}-{\bf r})=-C_6/|{\bf r}'-{\bf r}|^6$ ($C_6$ is the dispersion parameter).
%%%%%%%%%%%%%%%%%%%%%%
(b) Red solid circles show the atoms that have been excited to the Rydberg state $|4\rangle$.
Blue solid circles show the atoms that are not excited to the Rydberg state $|4\rangle$.
Large pale-red domains (with dashed circles) show Rydberg blockade spheres.
The large red arrow shows $\Omega_{c}$ (the half-Rabi frequency of the control field).
The $\mathcal{E}_{p+}$ and $\mathcal{E}_{p-}$ are the two circularly polarized components of the probe field $E_{p}$, $B$ is the weak static magnetic field (applied along the $z$ direction), and
$\psi$ is the deflection angle of the polarization vector of the probe field, resulted from the magneto-optical effect.}
\label{Fig1}
\end{figure}

We start by considering a cold, lifetime-broadened, inverted-Y-type four-level alkali (e.g. rubidium) atomic gas-mental, with the level diagram and excitation scheme shown in Fig.~\ref{Fig1}(a)~\cite{Sinclair2019,YanD2012}.
%===========================fig1===============================%
%===========================fig1===============================%
Here a weak probe laser field with two orthogonal, circularly polarized components (half Rabi frequencies are  $\Omega_{p1}$ and $\Omega_{p2}$, respectively) drives the transitions $|1\rangle\leftrightarrow |3 \rangle$ and $|2\rangle \leftrightarrow |3\rangle$, respectively, and a strong, linearly polarized control laser field with half Rabi frequency $\Omega_{c}$ drives the transition $|3\rangle\leftrightarrow|4\rangle$. The lower levels $|1\rangle$ and $|2\rangle$ are two Zeeman sublevels (split from a hyperfine ground-state level by a weak static magnetic field $B$ applied along the $z$ direction) and $|4\rangle$ is a high-lying Rydberg state with a large principal quantum number $n$. In the figure, $\Delta_3$ is one-photon detuning and $\Delta_2$
and $\Delta_4$ are two-photon detunings; $\Gamma_{jl}$ is the spontaneous-emission decay rate from $|l\rangle$ to $|j\rangle$; $\Gamma_{12}$ ($\Gamma_{21}$) represents the incoherent population transfer from $|1\rangle$ to $|2\rangle$ ($|2\rangle$ to $|1\rangle$). The atomic gas is laser cooled to an ultralow temperature and the probe (control) field propagates along the $z$ ($-z$) direction so that the first-order Doppler effect can be suppressed~\cite{Budker2002,Budker2007,Hang2007,Hang2012}. A schematic of the experimental geometry of the system is given in Fig.~\ref{Fig1}(b). Notice that in the inverted-Y-type excitation scheme illustrated in Fig.~\ref{Fig1}(a), there are two ladder-type excitation paths, i.e., $|1\rangle \rightarrow|3\rangle \rightarrow|4\rangle$ and $|2\rangle \rightarrow|3\rangle \rightarrow|4\rangle$, which constitute two standard Rydberg EITs (i.e., double Rydberg EIT, with the state $|4\rangle$ a shared Rydberg state).

The expression of the electric field in the system can be written in the form
$\mathbf{E}=\mathbf{E}_p+\mathbf{E}_c=(\hat{\epsilon}_{+} \mathcal{E}_{p+}+\hat{\epsilon}_{-} \mathcal{E}_{p-})\exp [i(k_{p}z-\omega_{p} t)]+\hat{\epsilon}_{c} \mathcal{E}_{c}\exp [i(-k_{c}z-\omega_{c} t)]+\mathrm{c.c.}$
%%%%%%%%%%%%%%%%%%%%%%%%
Here $\mathcal{E}_{p+}$ and $\hat{\epsilon}_{+}\equiv (\hat{\mathbf{x}}+i\hat{\mathbf{y}}) /\sqrt{2}$~[$\mathcal{E}_{p-}$ and $\hat{\epsilon}_{-}\equiv(\hat{\mathbf{x}}-i\hat{\mathbf{y}}) /\sqrt{2}$] are, respectively, the amplitude and unit vector of the right-circular polarization ($\sigma^+$) component [left-circular polarization ($\sigma^-$) component] of the probe field, with $\hat{\mathbf{x}}\,(\hat{\mathbf{y}})$ the unit vector along the $x\,(y)$ direction; $\mathcal{E}_{c}$ and
$\hat{\epsilon}_{c}$ are the amplitude and polarization unit vector of the control field, respectively;
$\omega_p$ and $k_p=\omega_p/c$ ($\omega_c$ and $k_c=\omega_c/c$) are the angular frequency and wavenumber of the probe (control) field, respectively. The dynamics of the system is controlled by the Hamiltonian $\hat{H}=\mathcal{N}_a\int d^3{ r}\hat{\mathcal{H}}({\bf r},t)$, where $\hat{\mathcal{H}}({\bf r},t)$ is the Hamiltonian density and $\mathcal{N}_a$ is the atomic density. Under the electric-dipole approximation and rotating-wave approximation (RWA) , the Hamiltonian density reads
%%%%%%%%%%%%%%%%%%%%%%%%%%%%%%%%%%
\begin{align}\label{eqn1}
\hat{\mathcal{H}}=&-\sum_{\alpha=2}^{4}\hbar \Delta_{\alpha}
\hat{S}_{\alpha\alpha}( {\bf r},t)\notag\\
&-\!\hbar\left[\Omega_{p1}^{\ast }\hat{S}_{31}\!({\bf r},t)\!+\!\Omega_{p2}^{\ast }
\hat{S}_{32}( {\bf r},t)\!+\!\Omega_{c}^{\ast }\hat {S}_{43}({\bf r},t)\!+\!{\rm H.c}.\right]\notag\\
&+\mathcal{N}_{a}\!\int\! d^3 r' \hat {S}_{44}({\bf r}',t) \hbar V({\bf r}^{\prime}-{\bf r})\hat{S}_{44}({\bf r},t),
\end{align}
%%%%%%%%%%%%%%%%%%%%%%%%%%%%%
%%%%%%%%%%%%%%%%%%%%%%%%%%%%%
where $d^3 r'=dx' dy' dz'$; H.c. represents the Hermitian conjugate;
$\Delta_2=\omega_{p1}-\omega_{p2}-(E_2-E_1)/\hbar=-(E_2-E_1)/\hbar$ (because $\omega_{p1}=\omega_{p2}=\omega_{p}$),
$\Delta_3=\omega_{p}-(E_3-E_1)/\hbar$, and $\Delta_4=\omega_c+\omega_p-(E_4-E_1)/\hbar$ are the detunings
(with $E_{\alpha} = \hbar \omega_{\alpha}$ the eigenenergy of the state $|\alpha\rangle$); and $\hat {S}_{\alpha \beta }=|\beta\rangle\langle\alpha|\exp\{i[( {\bf k}_{\beta}-{\bf k}_{\alpha})\cdot{\bf r}-(\omega_{\beta}-\omega _{\alpha }+\Delta_{\beta}-\Delta_{\alpha }) t]\}$~($\alpha,\,\beta$=1-4)
%%%%%%%%%%%%%%%%%%%%%%%%%
is the transition operator satisfying the commutation relation
%%%%%%%%%%%%%%%%%%%%%%%%%%%%%%%%%%%%%%%%%%%%%
\begin{eqnarray}
&& [\hat{S}_{\alpha\beta}({\bf
r},t),\hat{S}_{\alpha^\prime\beta^\prime}({\bf r}^\prime,t)]\nonumber\\
&& =\mathcal{N}_a^{-1}
\delta ({\bf r}-{\bf r}{^\prime})[\delta_{\alpha\beta'}\hat{S}_{\alpha'\beta}({\bf r},t)-\delta_{\alpha'\beta}\hat{S}_{\alpha\beta'}({\bf r'},t)].
\end{eqnarray}
%%%%%%%%%%%%%%%%%%%%%%
The half Rabi frequencies of the probe and control fields are defined by $\Omega_{p1}=( {\bf \bm{p}}_{13}\cdot\hat{\epsilon}_{-}){\cal
E}_{p-}/\hbar$ and $\Omega_{p2}=( {\bf \bm{p}}_{23}\cdot\hat{\epsilon}_{+}){\cal E}_{p+}/\hbar$ and by $\Omega_c=({\bf e}_{c}\cdot{\bf p}_{43}){\cal
E}_{c}/\hbar$, respectively, with ${\bf p}_{\alpha\beta}$  the electric dipole matrix element associated with the transition between the states $|\alpha \rangle$ and $|\beta\rangle$. The third line in the Hamiltonian density (\ref{eqn1}) is contributed by the Rydberg-Rydberg interaction, with the van der Waals potential of the form  $V({\bf r}^{\prime}-{\bf r})=-C_6/|{\bf r}^{\prime}-{\bf r}|^6$ ($C_6$ is the van der Waals dispersion parameter) describing the interaction between the atoms located at positions $\bf{r}$ and $\bf{r}'$, respectively~\cite{Pritchard2013-1,Firstenberg2016-1,Murray2016-1}. The Rydberg-Rydberg interaction results in energy shifts and hence induces a phenomenon called Rydberg blockade~\cite{Saffman2010,Adams2020}, by which only one atom can be excited to Rydberg states in any spatial region of the atomic ensemble (i.e. Rydberg blockade sphere) with radius $R_{b}$.
\footnote{Here $R_b$ can be estimated by the formula $R_b=(|C_6/\delta_{\rm{EIT}}|)^{1/6}$, where $\delta_{\rm{EIT}}$ is the linewidth of the EIT transmission spectrum, given by $\delta_{\rm{EIT}}\approx |\Omega_c|^2 /|\Delta_3|$ for $\Delta_3\gg\gamma_{31}$ and $\delta_{\rm{EIT}}=|\Omega_c|^2/\gamma_{31}$ for $|\Delta_3|=0$.} The detailed derivation of the Hamiltonian (\ref{eqn1}) is given in Appendix~\ref{appendix0}.

As indicated above, the levels $|1\rangle$ and $|2\rangle$ are originated from the Zeeman splitting of a hyperfine ground level. When the external magnetic field $B$ is applied, the level spacing between $|1\rangle$ and $|2\rangle$, i.e., $E_2-E_1$, equals $2\mu_{\rm{B}} g_{F}B$, which means that the two-photon detuning $\Delta_2=-2\mu_{\rm{B}}g_{F} B/\hbar$ (here $\mu_{\rm{B}}$ and $g_F$ are the Bohr magneton and gyromagnetic factor, respectively). One can speculate that the change of $B$ will produce changes in the atomic population and coherence and hence variations of the Kerr nonlinearity and polarization state of the probe field.
%%%%%%%%%%%%%%%%%%%%%%%%%

The dynamics of the atomic motion is controlled by the optical Bloch equation
\begin{align}\label{Bloch}
\frac{\partial \hat{\rho}}{\partial t}=-\frac{i}{\hbar}\left[\hat{H}, \hat{\rho}\right]-\Gamma\,[{\hat \rho}],
\end{align}
where $\hat{\rho}\equiv\langle\hat{S}\rangle$ is the one-body density matrix (DM) (with the DM elements  given by $\rho_{\alpha\beta}\equiv\langle\hat{S}_{\beta\alpha}\rangle$ for $\alpha, \beta$=1-4)
\footnote{Here $\langle \hat{S}_{\alpha\beta}\rangle=\langle \Psi(0)|\hat{S}_{\alpha\beta}|\Psi(0)\rangle$, with $|\Psi(0)\rangle$ the initial quantum state of the atomic gas where all the atoms are populated in the ground state $|1\rangle$ and $|2\rangle$ before the two probe fields are applied.} and $\Gamma$ is the relaxation matrix describing the spontaneous emission and dephasing. The explicit expression of Eq.~(\ref{Bloch}) is given by Eq.~(\ref{eqn2}) in the Appendix~\ref{appendixA}.

The evolution of the two polarization components of the probe field is described by the Maxwell equation $\nabla^2 {\bf E}_p-(1/c^2)\partial^2 {\bf E}_p/\partial t^2
=[1/(\varepsilon_0 c^2)]\partial^2 {\bf P}_p/\partial t^2$, with the polarization intensity defined by ${\bf P}_p=
{\cal N}_a \{({\bf p}_{13}\rho_{31}+{\bf p}_{23}\rho_{32})\exp [i({ k}_{p} z-\omega_{p} t)]+{\rm c.c.}\}$, where ${\bf \bm{p}}_{13}$ (${\bf \bm{p}}_{23}$) is the electric-dipole matrix element related to the transition from $|3\rangle$ to $|1\rangle$ ($|2\rangle$).
We assume that the photon number in the probe field is high so that a semi-classical description for the system can be adopted~\cite{Sevincli2011a,Stanojevic2013,Bai2016,
Tebben2019,Bai2019}. Under the paraxial and slowly varying envelope approximations, the Maxwell equation is reduced to
%%%%%%%%%%%%%%%%%%%%%%%%%%%%%%%%
\begin{subequations}\label{eqn3}
\begin{align}
&i\left(\frac{\partial}{\partial z} + \frac{1}{c}\frac{\partial} {\partial t} \right) \Omega_{{p}1}+\kappa_{13}\rho_{31}=0, \\
& i\left(\frac{\partial}{\partial z}+\frac{1}{c}\frac{\partial}{\partial t}\right) \Omega_{{p}2}+\kappa_{23} \rho_{32}= 0,
\end{align}
\end{subequations}
%%%%%%%%%%%%%%%%%%%%%%%%%%%%%%
where $\kappa_{13}=\mathcal{N}_{a}| {\bf \bm{p}}_{13}\cdot\hat{\epsilon}_{-}|^{2}
\omega_{{p}}/(2\varepsilon_{0}c\hbar)$ and $\kappa_{23}=\mathcal{N} _{a}| {\bf \bm{p}}_{23}\cdot\hat{\epsilon}_{+}|^{2}
\omega_{{p}}/(2\varepsilon_{0}c\hbar)$ are coupling constants,
with $\varepsilon_0$ the vacuum dielectric constant and $c$ the light speed in vacuum. Note that for simplicity, we have assumed that the probe field has a large beam radius in transverse (i.e. $x$ and $y$) directions,
so that the diffraction effect is negligible.
%%%%%%%%%%%%%%%%%%%%%%%%%%%%%%%%%

The physical model described above is principally valid for many alkali atomic gases, such as $^{85}$Rb, $^{87}$Rb and $^{87}$Sr. For comparison with the experimental result reported in Ref.~\cite{Sinclair2019}, in later numerical calculations we take cold $^{85}$Rb gas as an example, for which the atoms have total nuclear angular momentum $I=5/2$. The levels for realizing the double Rydberg EIT are selected to be
$|1\rangle=|5^{2} S_{1/2},F=3,m_{F}=1\rangle$, $|2\rangle=|5^{2} S_{1/2}, F=3, m_{F}=-1\rangle$, $|3\rangle=|5^{2} P_{3/2}, F=4, m_{F}=0\rangle$, and $|4\rangle=|68S_{1/2}\rangle$~\cite{data}.
The van der Waals dispersion parameter reads $C_6 \simeq-2\pi\times 625.6$ $\mathrm{GHz}\,\mu{\rm m}^6$~\cite{Singer2005} and the decay rates are given by
$\Gamma_{12}=\Gamma_{21}=2\pi\times0.0016\,\rm{MHz}$,
$\Gamma_3=2\pi\times6.06\,\rm{MHz}$,
$\Gamma_{4}=2\pi\times0.02\,\rm{MHz}$, and $\Gamma_{13}=\Gamma_{23}=\Gamma_3/2$.
For $^{87}$Rb atoms, though having different total nuclear angular momentum ($I=3/2$), one can choose similar
levels to realize the double Rydberg EIT and hence to implement related experiments.
%%%%%%%%%%%%%
Note that in our scheme the states $|3\rangle$ and $|4\rangle$ are not sensitive to the magnetic field $B$ because of the special choice of their magnetic quantum number, by which an effective coupling between the (linearly polarized) control field and the states $|3\rangle$ and $|4\rangle$ is allowed only for $m_F=0$.

%{\color{magenta}\large \it [Note: My revision is upto here].}

\subsection{Linear dispersion relation of the double EIT}

We first discuss the linear optical property of the double Rydberg EIT.
In a linear approximation, the Maxwell-Bloch equations.~(\ref{Bloch}) and (\ref{eqn3}) admit
the solution $\Omega_{{p}j}=F_j\exp(i\theta_j)$. Here $F_j$ is a constant, $\theta_{j}=K_{j}(\omega)z-\omega t$ ($j=1,2$),
\footnote{Generally, in the atomic gas the frequency and wave number of the probe field are given by $\omega_{p}+\omega$ and $k_p+K(\omega)$, respectively ($\omega_{p}$ and $\omega$ are the center and sideband frequencies, respectively). Thus the case $\omega= 0$ corresponds to the probe-field frequency taking its center frequency $\omega_p$.}
and $K_{1}(\omega)$  [$K_{2}(\omega)$] is the linear dispersion relation for the $\sigma^+$ ($\sigma^-$) polarization component of the probe field given by
%%%%%%%%%%%%%%%%%%%%%%%%%%%%%%%%%%%%%%%
\begin{align}\label{eqn4}
K_{j}(\omega)=&\frac{\omega}{c}+
\frac{\kappa_{j3}(\omega+d_{4j})}{2[|\Omega_{{c}}|^2
-(\omega+d_{3j})(\omega+d_{4j})]},
\end{align}
where $d_{\alpha j}=\Delta_{\alpha}-\Delta_{j}+i\gamma_{\alpha j}$  ($\alpha=3,4$ and $j=1,2$) are constants, with $\gamma_{\alpha j}$ the damping parameters related to the spontaneous emission and dephasing of the atoms (see the Appendix~\ref{appendixA}).

Figure.~\ref{Fig2} illustrates
%===========================fig2===============================%
\begin{figure}[htp]
\centering
\includegraphics[width=1\linewidth]{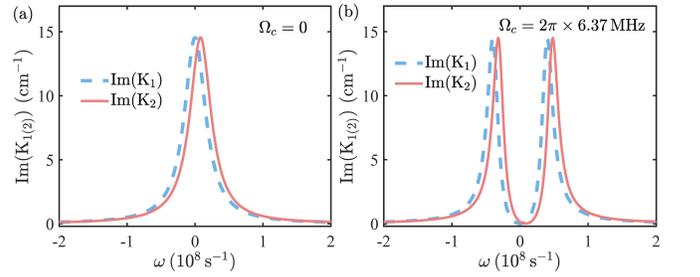}
\caption{\footnotesize Linear absorption spectra Im$(K_{1})$ (blue dashed line) and Im$(K_{2})$ (red solid line) for the $\sigma^+$ and $\sigma^-$ polarization components of the probe field, respectively, plotted
as functions of $\omega$ for $\Delta_2=2\pi\times 1.28\,\mathrm{MHz}$.
%%%%%%%%%%
(a)~Case with $\Omega_{{c}}=0$ (no EIT).
(b)~Case with $\Omega_{{c}}=2\pi\times 6.37\,\mathrm{MHz}$, for which an EIT window appears for both Im$(K_{1})$ and Im$(K_{2})$ (i.e., a double EIT occurs). When $\Delta_2=0$, Im$(K_{1})$ nearly coincides with  Im$(K_{2})$ (i.e., they are almost degenerate).
}
\label{Fig2}
\end{figure}
%===========================fig2===============================%
absorption spectra Im$(K_{1})$ (blue dashed line) and Im$(K_{2})$ (red solid line) as functions of $\omega$. When plotting the figure, the system parameters were chosen to be those given in the preceding subsection, together with $|\textbf{p}_{13}|\simeq|\textbf{p}_{23}|=3.58\times10^{-27}$ C cm~\cite{data}, $\mathcal{N}_a=1\times10^{10}$ cm$^{-3}$, $\Delta_3=\Delta_4=0$, and $\rho_{11}^{(0)}=\rho_{22}^{(0)}=0.5$.
\footnote{Principally, one can prepare different initial population distributions $\rho_{11}^{(0)}$ and $\rho_{22}^{(0)}$ ($\rho_{11}^{(0)}+\rho_{22}^{(0)}=1$) in the ground states $|1\rangle$ and $|2\rangle$. For simplicity, here we consider only the case $\rho_{11}^{(0)}=\rho_{22}^{(0)}=0.5$.}
Figure~\ref{Fig2}(a) shows the case in the absence of the control field (i.e., $\Omega_{{c}}=0$), for which no EIT occurs [i.e., both Im$(K_{1})$ and Im$(K_{2})$ have a single absorption peak]. Figure~\ref{Fig2}(b) shows the case in the presence of the control field ($\Omega_{{c}}=2\pi\times 6.37\,\mathrm{MHz}$), for which an EIT transparency window appears near $\omega=0$ for both Im$(K_{1})$ and Im$(K_{2})$. This means that two EITs occur (or a double EIT occurs) in the system. Note that if $\Delta_2=0$ the curves of Im$(K_{1})$ and Im$(K_{2})$ nearly coincide with each other, which means that the two EITs are nearly degenerate.

%%%%%%%%%%%%%%%%%%%%%%%%%%%%%%%%%%%%%%%%%%%%%%%%%
\section{Giant nonlocal self- and cross-Kerr nonlinearities}\label{sec3}

\subsection{Calculation of nonlinear optical susceptibilities}\label{sec30}

We now consider how to calculate the third-order nonlinear optical susceptibilities for this double Rydberg EIT system. Note that the electric polarization intensity of the probe field can be written in the form
$\mathbf{P}_p=\varepsilon_{0} ({\bf{\hat{\epsilon}_{-}}} \mathcal{E}_{p-}\chi_{1}+{\bf{\hat{\epsilon}_{+}}}\mathcal{E}_{p+}\chi_{2})\exp [i(k_{p} z-\omega_{p} t)]+{\rm{c.c.}}$ (c.c. means complex conjugate), where $\chi_1$ and $\chi_2$ are the optical susceptibilities of the two polarization components, respectively given by
%%%%%%%%%%%%%%%%%%%%%%%%%%%%%%%%%%%%
\begin{subequations}\label{eqn9}
\begin{align}
&\chi_{\rm{1}}=\frac{\mathcal{N}_{a}\left({\bf{\hat{\epsilon}_{-}}} \cdot \mathbf{{\bf \bm{p}}}_{13}\right) \rho_{31}}{\varepsilon_{0} \mathcal{E}_{p-}},\\
&\chi_{\rm{2}}=\frac{\mathcal{N}_{a}\left({\bf{\hat{\epsilon}_{+}}} \cdot \mathbf{{\bf \bm{p}}}_{23}\right) \rho_{32}}{\varepsilon_{0} \mathcal{E}_{p+}}.
\end{align}
\end{subequations}
%%%%%%%%%%%%%%%%%%%%%%%%%%%%%%%%%%%%%%
To acquire $\chi_{\rm{1}}$ and $\chi_{\rm{2}}$, one must solve the Bloch equation.~(\ref{Bloch}) to get one-body DM elements $\rho_{31}$ and $\rho_{32}$, which however depend on the two-body DM elements $\rho_{\alpha\beta,\mu\nu}({\bf r^{\prime},\bf r},t)=\langle\hat{S}_{\alpha\beta}\left(\mathbf {r}^{\prime}, t\right)\hat{S}_{\mu\nu}( \mathbf{r},t)\rangle$
due to the Rydberg-Rydberg interaction [see the explicit expression given in Eq.~(\ref{eqn2})]. Thus one must solve the equations for the two-body DM elements simultaneously, which however involves three-body DM elements
$\rho_{\alpha\beta,\mu\nu,\gamma\delta}({\bf r}^{\prime\prime},{\bf r}^{\prime}, {\bf r},t)=\langle
\hat{S}_{\alpha\beta}({\bf r}^{\prime\prime},t) \hat{S}_{\mu\nu}({\bf r}',t)\rangle \hat{S}_{\gamma\delta}({\bf r},t)\rangle$ and so on. For example, the equation for the two-body DM element for $\rho_{44,41}$ reads
\begin{align}
&\Big[i\frac{\partial}{\partial t}+i\Gamma_{34}+d_{41}-V\big(\mathbf{r}'-\mathbf{r}\big)\Big] \rho_{44,41}\notag\\
&+\Omega_{c}\rho_{34,41} +\Omega_{c}\rho_{44,31}-\Omega_{p1}\rho_{44,43}-\Omega_{c}^*\rho_{43,41} \notag\\
&-\mathcal{N}_{a} \int  d^{3} r^{\prime \prime}\rho_{44,44,41}\left(\mathbf{r}'',\mathbf{r}^{\prime},\mathbf{r}, t\right) V\left(\mathbf{r}''-\mathbf{r}\right)=0,\label{55,51}
\end{align}\noindent
which involves the three-body DM element $\rho_{44,44,41}$.
As a result, one obtains a chain of infinite equations for $N$-body reduced DM elements ($N=1,2,3,.\,.\,.,\infty$) (i.e., BBGKY hierarchy~\cite{Bogoliubov1992}), which are coupled to each other. For the problem where the Kerr effect is large, an effective approach beyond the MFA for solving these reduced DM elements must be developed. In particular, a technique to truncate such a chain of equations must be adopted. Our method is to employ the  RDME~\cite{Mukamel1995,Schempp2010,Sevincli2011b}
%%%%%%%%%%%%%%%%%%%%%%%%%%%%%%%
\begin{align}\label{cluster}
&\rho_{\alpha\beta,\mu\nu,\gamma \delta}({\bf r}^{\prime\prime},{\bf r}^{\prime},{\bf r},t)\nonumber\\
&=\rho_{\alpha\beta} ({\bf r}^{\prime\prime},t)\rho_{\mu\nu,\gamma\delta} ({\bf r}^{\prime},{\bf r},t)
+\rho_{\alpha\beta,\mu\nu} ({\bf r}^{\prime\prime},{\bf r}^{\prime},t)\rho_{\gamma \delta}({\bf r},t)\nonumber\\
&+\rho_{\alpha\beta,\gamma \delta} ({\bf r}^{\prime\prime},{\bf r},t)\rho_{\mu\nu}({\bf r}^{\prime},t)
-2\rho_{\alpha\beta}({\bf r}^{\prime\prime},t) \rho_{\mu\nu}({\bf r}^{\prime},t) \rho_{\gamma \delta}({\bf r},t),
\end{align}
%%%%%%%%%%%%%%%%%%%%%%%%%%%%%
for the three-body DM elements appeared in the equations of the two-body DM elements. In this way, the chain of infinite equations can be truncated consistently and thus the problem can be reduced to solving the closed equations for the one- and two-body DM elements only~\cite{Bai2016,Bai2019}. We stress that in our approach the GSA~\cite{note001} is not used, in which the equations of motion of the diagonal DM elements are solved together with those of the non-diagonal ones. Thus the approach is valid for cases where the system possesses strong Kerr nonlinear effects.

Similar to Ref.~\cite{Bai2016}, based on the RDME and noting that the calculation of $\chi_{\rm{1}}$ and $\chi_{\rm{2}}$ can be accomplished by solving the Bloch equation.~(\ref{Bloch}) [or the closed equations for the one- and two-body DM elements {in a steady state (i.e. $\partial/\partial t=0$)}] through a perturbation expansion by taking $\Omega_{pj}$ as small quantities, we can get the solution of $\rho_{3j}$ up to the third order approximation, with the result given by
$\rho_{3j}\simeq a_{3j}^{(1)} \Omega_{{pj}}+\sum_{l}^{1,2}a_{3j,l}^{(3,\rm{loc})}|\Omega_{{pl}}|^{2} \Omega_{{pj}}+\int d^3r'\sum_{l}^{1,2}a_{3j,l}^{(3,\rm{nloc})}|\Omega_{{pl}}({\bf {r}}^{\prime})|^{2} \Omega_{{pj}}$ ($j=1,\,2$). As a result, we obtain the expression of the optical susceptibility of $j$th polarization component of the probe field
%%%%%%%%%%%%%%%%%%%%%%%
\begin{align}\label{eqn10}
\chi_{j}=
&\,\chi_{j}^{(1)}+\chi_{j1}^{(3,\rm{loc})}\left|\mathcal{E}_{p-}\right|^{2}+\chi_{j2}^{(3,\rm{loc})}
\left|\mathcal{E}_{{p+}}\right|^{2}\notag\\
&+\chi_{j1}^{(3,\rm{nloc})}\left|\mathcal{E}_{p-}\right|^{2}
+\chi_{j2}^{(3,\rm{nloc})}\left|\mathcal{E}_{{p+}}\right|^{2}.
\end{align}
%%%%%%%%%%%%%%%%%%%%%%%%%%%%%%%%%%%%%%%%%%%%
Here
$\chi_{j}^{(1)}=\frac{\mathcal{N}_{a}\left|\mathbf{p}_{j3}
\right|^{2}}{\varepsilon_{0} \hbar}\frac{d_{4j}}{2D_{{j}}}$
%%%%%%%%%%%%%%%%%%%%%%%%%%%%%%%%%%%%%%%%%%%%%
is the first-order (linear) susceptibility, while
%%%%%%%%%%%%%%%%%%%%%%%%%
\begin{subequations}\label{eqn11}
\begin{align}
&\chi_{jl}^{(3,{\rm loc})}=\frac{\mathcal{N}_{a}b_j}{D_{j}} \left[d_{4l}\left(a_{jj,l}^{(2)}
- a_{33,l}^{(2)}\right)+\Omega _{{c}}^{\ast}a_{43,l}^{(2)}+\delta_{jl}d_{4l}a_{lj}^{(2)}\right],\label{chilocal}\\
%%%%%%%%%%%%%%%%%%%%%%%%
&\chi_{jl}^{(3,\rm{nloc})}=\frac{\mathcal{N}_{a}^2b_j\Omega_c^*}{D_{j}}\,\int d^3r^{\prime} a_{44,4j,l}^{(3)}\left(\mathbf{r}^{\prime}-\mathbf{r}\right) V\left(\mathbf{r}^{\prime}-\mathbf{r}\right)
,\label{chinonlocal}
\end{align}
\end{subequations}
are the local and nonlocal third-order nonlinear susceptibilities, contributed by the photon-atom interaction~\cite{note000} and the Rydberg-Rydberg interaction [which is nonlocal, manifested by the integration in (\ref{chinonlocal})], respectively. In the expressions, $D_{{j}}=|\Omega_{c}|^2-d_{3{j}}d_{4{j}}$ and $b_j=\left|\mathbf{p}_{j3}\right|^{4}/(\varepsilon_{0} \hbar^{3})$.
Physically, the case for $j=l$  ($j\neq l$) comes from the self-Kerr  (cross-Kerr) effect of the system, describing the self-phase (cross-phase) modulation of the probe field.
The detailed derivation of the result (\ref{eqn10}) and explicit expressions of $a_{jj,l}^{(2)}$, $a_{43,l}^{(2)}$, $a_{lj}^{(2)}$, and $a_{44,4j,l}^{(3)}$ in the expressions (\ref{eqn11}) are presented in the Appendixes \ref{appendixB} and \ref{appendixB5}, respectively. A notable character of the nonlocal Kerr nonlinear susceptibilities $\chi_{jl}^{(3,{\rm nloc})}$ is that they are proportional to $\mathcal{N}_{a}^2$, while the local Kerr nonlinear susceptibilities  $\chi_{jl}^{(3,{\rm loc})}$ are proportional to $\mathcal{N}_{a}$.
Because the local nonlinear susceptibilities $\chi_{jl}^{(3,{\rm loc})}$ are generally three orders of magnitude smaller than the nonlocal nonlinear susceptibilities $\chi_{jl}^{(3,{\rm nloc})}$ (they become vanishing when the two-photon detuning $\Delta_4=0$), we will neglect them in the following discussions.

%%%%%%%%%%%%%%%%%%%%%%%%%%%%%%%
\subsection{Kerr nonlinearities in dispersion regime}\label{sec31}
%%%%%%%%%%%%%%%%%%%%%%%%%%%%%%%

Based on the analytical result given above, numerical values of the third-order Kerr nonlinear optical susceptibilities can be calculated for realistic system parameters. The Kerr nonlinearities may be divided into different regimes depending on the system parameters, mainly depending on the ratio between the single-photon detuning $\Delta_3$ and the decay rate $\Gamma_3$ ($=\Gamma_{13}+\Gamma_{23}$) of the intermediate state $|3\rangle$~\cite{Sevincli2011a,Bai2016,Pritchard2013-1,Firstenberg2016-1,Murray2016-1}. In particular, if $\Delta_3$ is much larger (smaller) than $\Gamma_3$, the Kerr nonlinearity is in a dispersion (dissipation) regime.

We first study the case of the dispersion regime by taking
$\Delta_3=2\pi\times 100\,\mathrm{MHz}$ (i.e. the ratio $\Delta_3/\Gamma_3=16.5\gg1$),
$\Delta_2=2\pi\times0.5\,\mathrm{MHz}$ (corresponding to $B=11.9\,\mu$T), {$\Delta_4=2\pi\times 2\,\mathrm{MHz}$, $\Omega_{{c}}=2\pi\times 31\,\mathrm{MHz}$,
$\mathcal{N}_a=4\times10^{10}\,\mathrm{cm^{-3}}$, and $\rho_{11}^{(0)}=\rho_{22}^{(0)}=0.5$.
Because of the large detunings, the dephasing rates $\gamma_{jl}^{\rm dep}$ play no significant role and thus can be neglected.
Based on the formula (\ref{chinonlocal}), we obtain the result of the self-Kerr and the cross-Kerr nonlinear optical susceptibilities, given in Table~\ref{TAB1}.
%%%%%%%%%%%%%%%%%%%%%%%%%%%%%%%%%%%%%%%
\begin{table}
\renewcommand\tabcolsep{17pt}
\centering
\caption{
\footnotesize Third-order Kerr nonlinear optical susceptibilities in the dispersion regime:  $\chi_{11}^{(3,{\rm nloc})}$ and $\chi_{22}^{(3,{\rm nloc})}$ are the third-order self-Kerr nonlinear susceptibilities of the first and second polarization components of the probe field, respectively, and $\chi_{12}^{(3,{\rm nloc})}$ and $\chi_{21}^{(3,{\rm nloc})}$ are third-order cross-Kerr nonlinear susceptibilities between the two polarization components of the probe field.
The system parameters are
$\Delta_2=2\pi\times0.5\,\mathrm{MHz}$ (i.e., ${B}=11.9$ ${\rm{\mu T}}$), $\Delta_3=2\pi\times100$ $\mathrm{MHz}$,
$\Delta_4=2\pi\times2$ $\mathrm{MHz}$, $\Omega_{{c}}=2\pi\times31$ $\mathrm{MHz}$, and $\mathcal{N}_a=4\times10^{10}$ $\mathrm{cm^{-3}}$.
}
%%%%%%%%%%%%%%%%%%%%%%%%%%%%%%%%%%%
\vspace{0.2cm}
\label{TAB1}
%\begin{ruledtabular}
\begin{tabular}{cc}
\hline\hline\vspace{-0.3cm}&\\
Susceptibility&
Value$\,(\mathrm{m}^{2}$ $\mathrm{V}^{-2})$\\
\hline
&\vspace{-0.2cm}\\
%%%%%%%%%%%%%%%%%%%%%
$\chi_{11}^{(3,\rm{nloc})}$
&$(-1.5107+i0.0063)\times10^{-8}$\\
%%%%%%%%%%%%%%%%%%%%%
$\chi_{12}^{(3,\rm{nloc})}$
&$(-1.3519+i0.0049)\times10^{-8}$\\
%%%%%%%%%%%%%%%%%%%%%
$\chi_{21}^{(3,\rm{nloc})}$
&$(-1.3521+i0.0115 )\times10^{-8}$\\
%%%%%%%%%%%%%%%%%%%%%
$\chi_{22}^{(3,\rm{nloc})}$
&$(-1.2092+i0.0097 )\times10^{-8}$
%%%%%%%%%%%%%%%%%%%%%
\\
\hline\hline
\end{tabular}
%\end{ruledtabular}
\end{table}
%%%%%%%%%%%%%%%%%%%

From the table we see that the Kerr nonlinear susceptibilities $\chi_{jl}^{(3,\rm{nloc})}$  possess the following interesting features.
%%%%%%%%%%%%%%%%%%%%%%

(i) The values of the real parts of both the self-Kerr and cross-Kerr nonlinear susceptibilities can reach the order of magnitude $10^{-8}\,\mathrm{m}^2\,\mathrm{V}^{-2}$ for atomic density $\mathcal{N}_a=4\times10^{10}\,\mathrm{cm^{-3}}$. Such giant Kerr nonlinearities stem from the strong Rydberg-Rydberg interaction in the atomic ensemble.

(ii) The imaginary parts of the self-Kerr (cross-Kerr) nonlinear susceptibilities (i.e.,
${\rm{Im}}[\chi_{jl}^{(3,\rm{nloc})}]$) are much smaller than the corresponding real parts (i.e., ${\rm{Re}}[\chi_{jl}^{(3,\rm{nloc})}]$). Thus the nonlinear absorption of the probe field can be largely suppressed, which is due to the EIT effect and the contribution of the large detuning $\Delta_3$.

(iii) The self-Kerr nonlinear susceptibilities $\chi_{jj}^{(3,\rm{nloc})}$ ($j=1,2$) have the same orders as the cross-Kerr nonlinear susceptibilities $\chi_{jl}^{(3,\rm{nloc})}$ ($j,l=1,2 and j\neq l$) even for a low atomic density. This is quite different from those obtained by conventional EIT systems (without the Rydberg-Rydberg interaction) where the self-Kerr nonlinear susceptibilities can be made much smaller than the cross-Kerr nonlinear susceptibilities for a low atomic density~\cite{Petrosyan2004}.

The giant Kerr nonlinear susceptibilities obtained in the present double Rydberg EIT system are very promising for realizing many nonlinear and quantum optical processes, such as quantum phase gates, few-photon bound states, and quantum nondemolition measurements~\cite{Pritchard2013-1,Firstenberg2016-1,Murray2016-1}, and also for realizing nonlocal weak-light vector solitons and vortices, and so on.

%%%%%%%%%%%%%%%%%%%%%%%%%%%%%%%%%
\subsection{Kerr nonlinearities in dissipation regime}\label{sec32}

We turn to study the case of the dissipation regime of the Kerr nonlinearity, which can be realized by taking smaller values of $\Delta_3$. As an example, we take $\Delta_3=0$, and the other system parameters are
%%%%%%%%%%%%%%%%%%%%%%%%%%%%%%%%%%%%%%%
$\Delta_2=0$ (i.e., $B=0$), $\Delta_4=0$, $\Omega_{{c}}=2\pi\times9\,\mathrm{MHz}$, $\mathcal{N}_a=4\times10^{10}\,\mathrm{cm^{-3}}$, and $\rho_{11}^{(0)}=\rho_{22}^{(0)}=0.5$.
%%%%%%%%%%%%%%%%%%
In this situation, some additional decoherence processes (such as high-order Doppler effects,
\footnote{The high-order (residual) Doppler effects in the Rydberg EIT are much larger than that in conventional EIT because the probe and control fields in the system of Rydberg EIT have a large wavelength difference.}
dephasing of the Rydberg state by decaying to nearby $nP$ states, blackbody radiation, atomic collisions, and frequency instability of the lasers used) will be significant and hence should be taken into account in the theoretical calculation. To effectively include these decoherence processes, we assume
{$\gamma_{31}^{\rm dep}=\gamma_{32}^{\rm dep}=2\pi\times90\,\mathrm{MHz}$, and
%%%%%%%%%%%%%%%%%%
$\gamma_{41}^{\rm dep}=\gamma_{42}^{\rm dep}=2\pi\times1.8\,\mathrm{MHz}$.}
%%%%%%%%%%%%%%%%%%%%%%%%%
Due to the complete resonance, one maybe expects that the Kerr effect in this dissipation regime could be larger than that obtained in  the dispersion regime given in the preceding section, but the additional coherence processes will play roles in lowering the Kerr nonlinearities of the system.

With the use of the formula (\ref{chinonlocal}), the self-Kerr and the cross-Kerr nonlinear optical susceptibilities in the dissipation regime can be numerically calculated, with the result presented in Table~\ref{TAB2}.
%===========================TAB2===============================%
\begin{table}
\centering
\caption{\footnotesize
Third-order Kerr nonlinear optical susceptibilities in the dissipation regime: $\chi_{11}^{(3,{\rm nloc})}$ and $\chi_{22}^{(3,{\rm nloc})}$ are the third-order self-Kerr nonlinear susceptibilities of the first and second polarization components of the probe field, respectively, and $\chi_{12}^{(3,{\rm nloc})}$ and $\chi_{21}^{(3,{\rm nloc})}$ are third-order cross-Kerr nonlinear susceptibilities between the two polarization components of the probe field.
%============parameters===================%
The system parameters are
$\Delta_2=0$ (i.e., $B=0$), $\Delta_3=\Delta_4=0$, $\Omega_{{c}}=2\pi\times9$ $\mathrm{MHz}$, and $\mathcal{N}_a=4\times10^{10}$ $\mathrm{cm^{-3}}$.}
%============parameters===================%
%===========explanation===================%
\vspace{0.2cm}
\label{TAB2}
\renewcommand\tabcolsep{17pt}
%\begin{ruledtabular}
\begin{tabular}{cc}
\hline\hline\vspace{-0.3cm}&\\
Susceptibility &Value$\,(\mathrm{m}^{2}$ $\mathrm{V}^{-2})$\\
\hline
&\vspace{-0.2cm}\\
%%%%%%%%%%%%%%%%%
$\chi_{11}^{(3,\rm{nloc})}$
&$\,\,\,\,\,(
-1.4946+i1.4948)\times10^{-8}$\\
%%%%%%%%%%%%%%%%%%%%%%%%
$\chi_{12}^{(3,\rm{nloc})}$
&$\,\,\,\,\,(-1.4946+i1.4948)\times10^{-8}$\\
%%%%%%%%%%%%%%%%%%%%%%%%%
$\chi_{21}^{(3,\rm{nloc})}$
&$\,\,\,\,\,(-1.4946+i1.4948)\times10^{-8}$\\
%%%%%%%%%%%%%%%%%%%%%%%%
$\chi_{22}^{(3,\rm{nloc})}$
&$\,\,\,\,\,(-1.4946+i1.4948)\times10^{-8}$\\
\hline\hline
\end{tabular}
%\end{ruledtabular}
\end{table}
%===========================TAB2===============================%.
From the table, we see that the Kerr nonlinear susceptibilities $\chi_{jl}^{(3,\rm{nloc})}$ in this regime have the following characters:
%%%%%%%%%%%%%%%%%%%%%

(i) The values of the self-Kerr and cross-Kerr nonlinear susceptibilities can reach the order of $10^{-8}\,\mathrm{m}^2\,\mathrm{V}^{-2}$ for atomic density $\mathcal{N}_a=4\times10^{10}$ $\mathrm{cm^{-3}}$, which also stems from the strong Rydberg-Rydberg interaction.

(ii) Different from the dispersion regime, the imaginary parts of the Kerr nonlinear susceptibilities ${\rm Im}[\chi_{jl}^{(3,\rm{nloc})}]$ (for both the self-Kerr and cross-Kerr ones) in this dissipation regime have the same orders of magnitude as their corresponding real parts ${\rm Re}[\chi_{jl}^{(3,\rm{nloc})}]$, which means that both the nonlinear optical absorption and the phase modulation are significant in the system.

%%%%%%%%%%%%%%%%%%%%%%%%%%%%

Recently, the first experimental measurement on the real part of the cross-Kerr nonlinear susceptibility of a cold $^{85}$Rb gas was reported in Ref.~\cite{Sinclair2019}, where the system works in a dissipation regime with the parameters $\Delta_j$ ($j=2,3,4$), $\Omega_{{c}}$, and $\mathcal{N}_a$ the same as those given in caption of the Table~\ref{TAB2}. We note that there is a small difference between our theoretical result given here and the experimental one (i.e. ${\rm Re}[\chi_{12}^{(3,\rm{nloc})}]\approx  1.0\times 10^{-8}$ m$^2$\,V$^{-2}$~\footnote{The real parts of the third-order Kerr nonlinear susceptibilities (i.e., Re[$\chi^{(3,\rm{nloc})}_{jl}]$) are negative (self-defocusing type) because the Rydberg-Rydberg interaction in the system is repulsive.}) reported in Ref.~\cite{Sinclair2019}. The physical reason for this  difference is probably due to other physical factors (e.g., the involvement of multiple Rydberg states, imperfect EIT, or some other unknown noise), which existed in the experiment~\cite{Sinclair2019} but are not considered in our theoretical approach. Clarifying this difference is a topic deserving further exploration.

Generally, one can tune the ratio $\Delta_3/\Gamma_3$ to realize a transition from the dispersive Kerr nonlinearity to the dissipative Kerr nonlinearity. In addition, the values of the Kerr nonlinear susceptibilities can be further optimized by the choice of system parameters, which are omitted here. The giant Kerr nonlinearities in the dissipation regime have promising applications for generating single photons and realizing all-optical switches and transistors at single-photon levels, etc.~\cite{Pritchard2013-1,Firstenberg2016-1,Murray2016-1}.

\subsection{Comparison between the results obtained by MFA, GSA, and RDME}\label{sec33}

For calculations of the Kerr nonlinearities in Rydberg atomic gases, there exist three theoretical approaches: the MFA, RDME, RDME with GSA (here called as GSA for simplicity)~\cite{Sevincli2011a,Stanojevic2013,Grankin2015,Boddeda2016,Bienias2016,Bai2016,
Tebben2019,Bai2019,Schempp2010,Tong2004,Weimer2008,Petrosyan2011,Carr2013,Yan2013}. The differences between these approaches are in the methods of treating high-order many-body correlators (i.e., many-body DM elements) that appear in the equations of the lower-order correlators. For completeness here we compare on the self- and cross-Kerr nonlinear optical susceptibilities for our model obtained through the MFA, GSA, and RDME, respectively.

%%%%%%%%%%%%%%%%%%%%%%%%%%%%%%%%%%%%%
(i) {\it The MFA}. In this approach, one assumes that any two-body correlators can be decomposed as
\begin{equation}\label{approximationa}
\rho_{\alpha\beta,\mu\nu}(\mathbf {r}^{\prime},\mathbf{r},t)
\approx \rho_{\alpha\beta}(\mathbf{r}^{\prime}, t)\rho_{\mu\nu}(\mathbf{r},t).
\end{equation}
With such an assumption, the one-body DM equation (\ref{Bloch}) becomes self-closed and hence can be solved very simply. However, in this approach the atom-atom correlations are completely neglected, which is not valid for systems with strong atom-atom interaction. In particular, it is broken for many nonlinear optical processes in Rydberg gases due to the strong Rydberg-Rydberg interaction~\cite{Schempp2010}.
Under the MFA and using the same system parameters as those given in Sec.~\ref{sec31}, one obtains the Kerr nonlinear susceptibilities of the system, given by
$\chi^{(3,\rm{nloc})}_{jl}\simeq-(1.18+i 0.02)\times10^{-2}\,\mathrm{m}^{2}\,\mathrm{V}^{-2}$,
which is obviously unacceptable. Nevertheless, the MFA becomes valid if the atomic density ${\cal N}_a$ is very low so that the correlation effect between atoms plays no significant role.

%%%%%%%%%%%%%%%%%%%%%%%%%%%%%%%%%%%%%
(ii) {\it The GSA}. In this approach, one assumes that the diagonal elements of the one-body DM (i.e., the population in the atomic quantum states) are not changed during the time evolution of the system, e.g., $\rho_{11}=a_0$ and
$\rho_{22}=b_0$ (with $a_0+b_0=1$) and $\rho_{33}=\rho_{44}=0$. With the GSA, the number of the equations for the one-body DM elements is largely reduced, which also results in a significant reduction of the number of equations for the high-order DM elements.~\footnote{In the GSA, there is another assumption that must be used for decomposing the two-body DM elements (which describe the Rydberg-Rydberg interaction appearing in the equations of the one-body DM elements) into three-body DM elements. In order to make the problem closed and solvable, a truncation of infinite hierarchy of equations is still necessary.} In this way, the calculation of the nonlinear optical evolution of the system can be simplified greatly. However, such an approximation is questionable for Rydberg gases because the change in the atomic population cannot be neglected due to the strong Rydberg-Rydberg interaction, especially for the case with a high atomic density. Hence one must take into account and solve the equations of motion for the diagonal DM elements, which give non-zero $\rho_{33}$ and $\rho_{44}$ and hence non-negligible contributions to the Kerr nonlinear optical susceptibilities in the system~\cite{Bai2016}.

%%%%%%%%%%%%%%%%%%%%%%%%%%%%%%%%%%%%%%%%%%%%%%

(iii){\it The RDME}.
This is the approach used in our calculations~\cite{Bai2016,Bai2019}, in which all the equations for the diagonal and nondiagonal one-body DM elements are considered; the three-body DM elements appearing in the equations of the two-body DM elements are factorized by using the formula (\ref{cluster}) and hence the equations for the one- and two-body DM become closed. It is an approach beyond both the MFA and the GSA, and can be solved
by using the perturbation expansion, presented in Appendix \ref{appendixB5}. Note that the contribution by the two-body DM elements starts only from the second-order approximation; for the inverted-Y-type system considered here, in the third-order approximation of the perturbation expansion there are eight correlator equations for each component of the probe field, given by Eq.~(\ref{third-order two-body}).

For a comparison of the three theoretical approaches, Fig.~\ref{Fig3}
%===========================fig3===============================%
\begin{figure}
\centering
\includegraphics[width=1\linewidth]{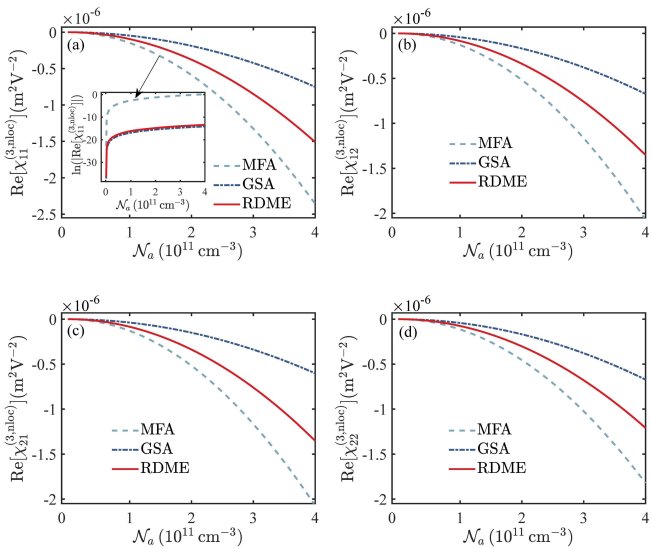}
\caption{\footnotesize Comparison between the results of the Kerr nonlinear optical susceptibilities $\chi_{jl}^{(3,\rm{nloc})}$ ($j,l=1,2$) in the dispersion regime, obtained by using the MFA, GSA, and RDME.
%%%%%%%%%%%%%%%%%%%%%%%%%%%%%%%%%%%%%%%%%%%%
(a)~Real part of $\chi_{11}^{(3,\rm{nloc})}$ (i.e., Re[$\chi_{11}^{(3,\rm{nloc})}$]) as a function of atomic density $ \mathcal{N}_a$, obtained by the MFA (gray dashed line), GSA (blue dash-dotted line), and RDME (red solid line).
%%%%%%%%%%%%%%%%%%%%%%%%%%%%%%%%%%%%%%%
(b)–(d)~Similar to (a) but for (b) Re[$\chi_{12}^{(3,\rm{nloc})}$], (c) Re[$\chi_{21}^{(3,\rm{nloc})}$], and (d) Re[$\chi_{22}^{(3,\rm{nloc})}$].
The system parameters are
$\Delta_2=2\pi\times0.5$ $\mathrm{MHz}$,  $\Delta_3=2\pi\times100$ $\mathrm{MHz}$, $\Delta_4=2\pi\times2$ $\mathrm{MHz}$, and
$\Omega_{{c}}=2\pi\times31$ $\mathrm{MHz}$.
The susceptibilities obtained by the MFA shown in all four panels have been divided by $2\times10^{6}$. The inset in (a) is the natural logarithm diagram of $|{\rm Re}[\chi_{11}^{(3,\rm{nloc})}]|$ for the results obtained by the MFA, GSA, and RDME.
}
%============explanation===================%
\label{Fig3}
\end{figure}
%===========================fig3===============================%
shows the results of the real parts of the self-Kerr and cross-Kerr nonlinear optical susceptibilities $\chi_{jl}^{(3,\rm{nloc})}$,  i.e., Re[$\chi_{jl}^{(3,\rm{nloc})}$] ($j,l=1,2$) in a dispersion regime, which are obtained by exploiting the MFA (gray dashed lines), GSA (blue dash-dotted lines), and RDME (red solid lines). Illustrated are Re[$\chi_{11}^{(3,\rm{nloc})}$]~[Fig.\ref{Fig3}(a)], Re[$\chi_{12}^{(3,\rm{nloc})}$]~[Fig.\ref{Fig3}(b)], Re[$\chi_{21}^{(3,\rm{nloc})}$]~[Fig.\ref{Fig3}(c)], and Re[$\chi_{22}^{(3,\rm{nloc})}$]~[Fig.\ref{Fig3}(d)] as functions of the atomic density ${\cal N}_a$. When plotting the figure, the system parameters were chosen to be
$\Delta_2=2\pi\times 0.5\,\mathrm{MHz}$, $\Delta_3=2\pi\times 100\,\mathrm{MHz}$, $\Delta_3=2\pi\times 2\,\mathrm{MHz}$, and $\Omega_{{c}}=2\pi\times31\,\mathrm{MHz}$. Note that for the convenience of the comparison, the Kerr nonlinear susceptibilities obtained by the MFA shown in all four panels have been divided by $2\times10^{6}$. The inset in Fig.\ref{Fig3}(a) is the natural logarithm diagram of $|{\rm Re}[\chi_{11}^{(3,\rm{nloc})}]|$ for the results obtained by the MFA, GSA, and RDME, respectively.

By inspecting Fig.~\ref{Fig3}, we arrive at the following conclusions.
(i)~The Kerr nonlinear susceptibilities obtained by using the MFA are much larger than those obtained by using the GSA and RDME, which are not physically reasonable, in particular for the case of high atomic density.
(ii)~For low atomic density, the results obtained by the GSA and by the RDME are closed; however, for high atomic density the results obtained by these two approaches display very different behaviors. In particular, the values given by the GSA are lower than those given by the RDME. This means that the contributions by the nonzero $\rho_{33}$ and $\rho_{44}$ (disregarded in the GSA) cannot be neglected.

\section{Giant magneto-optical rotations}\label{sec4}

Finally, as one of the promising applications of the double Rydberg-EIT and the enhanced Kerr effects illustrated above, we consider the MOR~\cite{Budker2002} of the system and show that it is possible to realize a giant enhancement of the MOR of the probe field if an external magnetic field ${\bf B}=(0,0,B)$ is applied and the system works in the dispersive nonlinearity regime. The MOR enhancement can be used to design a magnetometer~\cite{Budker2007} that can measure very weak magnetic fields with very high precision.

As indicated in Sec.~\ref{sec2}\,A, when the magnetic field ${\bf B}$ is present the Zeeman effect generated by the magnetic field makes the levels $|1\rangle$ and $|2\rangle$ (which are degenerate when $B=0$) produce the energy spacing $\Delta{ E}=E_2-E_1=2\mu_{\rm{B}}g_{F}B$. For the $D2$ line of $^{85}$Rb atoms, $g_F=1/3$; thus we have the expression of the two-photon detuning [see Fig.~\ref{Fig1}(a)]
\begin{align}
\Delta_2 =-\frac{2\mu_B B}{3\hbar},
\end{align}
which will appear in Bloch equation.~(\ref{Bloch}) (see the explicit expression presented in the Appendix~\ref{appendixA} where $d_{\alpha\beta}\equiv \Delta_{\alpha}-\Delta_{\beta}+i\gamma_{\alpha\beta}$) and result in a MOR of the probe field.
We stress that, due to the choice of magnetic quantum numbers and the linear polarization of the control field, the levels $|3\rangle$ and $|4\rangle$ are not sensitive to the applied magnetic field.

To investigate the MOR, one needs the nonlinear coupled equations controlling the evolution of the envelopes of the two polarization components of the probe field. Such equations can be derived by using the method of multiple-scales~\cite{Newell1990,Bai2019}, and have the nondimensional form
%%%%%%%%%%%%%%%%%%%%%%%%%%%%%%%%%%%%%%%%%%%%%%%%%%%%%%
\begin{align}\label{eqn8}
i\frac{\partial{u_{j}}} {\partial s }&
-\left(\sum_{l=1}^2 w_{jl}\left|u_{l}\right|^{2}\right)u_{j} \notag\\
&-\int d^3\zeta'\sum_{l=1}^2 g_{jl}(\vec{\zeta}'-\vec{\zeta})|u_{l}(\vec{\zeta}^{\prime})| ^{2}
%+\!g_{jl}\left|u_{l}\left(\bf{r}^{\prime}\right) \right|^{2}\right)
u_{j}(\vec{\zeta})= 0,
\end{align}
%%%%%%%%%%%%%%%%%%%%%%%%%%%%%%%
where we have defined $u_{j}=F_j/U_0$ ($F_j$ is the envelope of the $j$th polarization component, with $j=1,2$),  $s=z/L_{\rm NL}$, and $\vec{\zeta}={\bf r}/R_0=(\xi,\eta,s)$.
In these definitions, $U_0$ is the typical half Rabi frequency, and $L_{\rm NL}=1/(|W_{22}|U_0^2)$ is the nonlinearity length. In Eq.~(\ref{eqn8}) we have also defined  $w_{jl}=W_{jl}/|W_{22}|$ and $ g_{jl}=G_{jl}/|W_{22}|$.
Here $W_{jl}$  (proportional to $\chi_{jl}^{(3,\rm{loc})}$) are coefficients of local Kerr nonlinearities  characterizing the self-phase modulations (SPMs) for $j=l$ and cross-phase modulations (CPMs) for $j\neq l$, and $G_{jl}$  (proportional to $\chi_{jl}^{(3,\rm{nloc})}$) are coefficients of nonlocal Kerr nonlinearities  characterizing the nonlocal SPMs for $j=l$ and CPMs for $j\neq l$. The detailed derivation of Eq.~(\ref{eqn8}) and
explicit expressions of $W_{jl}$ and $G_{jl}$ are presented in the Appendix~\ref{appd}.

Equation (\ref{eqn8}) admits the exact solution (describing phase modulation)  $u_j=A_j \exp\left(-i\varphi_{j}s\right)$, where $\varphi_j=\sum_{l=1}^2\left[w_{jl}+\int d^3\zeta g_{jl}(\vec{\zeta})\right]A_l^2$, with $A_j$  ($j=1,2$) arbitrary constants.
Expressed by the original variables, the solution takes the form
\begin{align}
  &\Omega_{pj}=U_0A_j\exp(iK_{j}L)\notag\\
  &\times\exp\left\{-\sum_{l=1}^2 A_l^2\left[w_{jl}+\int d^3\zeta g_{jl}(\vec{\zeta})\right]L/L_{\rm{NL}}\right\},
\end{align}
%%%%%%%%%%%%%%%%%%%
where $L$ is the length of the atomic cell. To illustrate how the MOR occurs, we define the parameters $\psi_{\rm{lin}}$, $\psi_{\rm{loc}}$, $\psi_{\rm{nloc}}$, and $\psi_{\rm{tot}}$, which denote the rotation angles contributed by the linear, local nonlinear, nonlocal nonlinear, and total magneto-optical effects in the system, respectively. A simple calculation gives the following expressions:
\begin{subequations}\label{T}
\begin{align}
&\psi_{\rm{lin}} =\frac{L}{2}\left(\varphi_1-\varphi_2\right), \\
&\psi_{\rm{loc}}= \frac{L}{2L_{\rm{NL}}}\left[(w_{11}+w_{12})A_1^2-(w_{21}+w_{22})A_2^2\right],
\label{psilocal} \\
&\psi_{\rm{nloc}} \!=\!\frac{L}{2 L_{\rm{NL}}}\!\int\! d^3\zeta\sum_{l=1}^2\left[ g_{1l}(\vec{\zeta})A_1^2\!- g_{2l}(\vec{\zeta})A_2^2\right].\label{psinonlocal}\\
&\psi_{\rm{tot}}=\psi_{\rm{lin}}-\psi_{\rm{loc}}-\psi_{\rm{nloc}}.
\label{psitotoal}
\end{align}
\end{subequations}
Figure~\ref{Fig4}(a)
%===========================fig4===============================%
\begin{figure}
\centering
\includegraphics[width=0.85\linewidth]{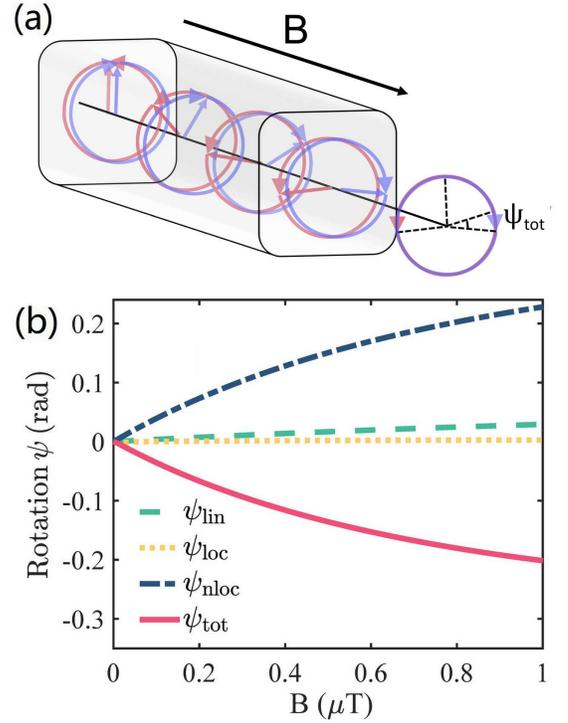}
\caption{{\footnotesize Giant magneto-optical rotation of the probe field contributed by the enhanced Kerr effects.
%%%%%%%%%%%%%%
(a)~Schematic of the MOR with the magnetic field applied along the propagation (i.e. $z$) direction of the probe field. The red (blue) solid circular curve with an arrow is the polarization direction for the circular polarization component $\mathcal{E}_{p+}$  ($\mathcal{E}_{p-}$), which is orthogonal to the propagation direction. In order to facilitate the distinction, the two components in the same position are separated by a little distance.
Here $\psi_{\rm{tot}}$ is the rotation angle due to the MOR.
(b)~The MOR angles as functions of the applied magnetic field $B$ for the atomic-cell length $L=5\,{\rm{mm}}$.
Here $\psi_{\rm{lin}}$ is the rotation angle due to the linear MOR (green dashed line),
$\psi_{\rm{loc}}$ is the rotation angle due to the local Kerr nonlinearities (yellow dotted line),
$\psi_{\rm{nloc}}$ is the rotation angle due to the nonlocal Kerr nonlinearities (blue dash-dotted line), and
$\psi_{\rm{tot}}=\psi_{\rm{lin}}-\psi_{\rm{loc}}-\psi_{\rm{nloc}}$ is the total rotation angle (red solid line). }}
\begin{comment}
Since $\psi_{\rm{lin}}$ and $\psi_{\rm{loc}}$ are nearly overlapped, for distinction the curve of $\psi_{\rm{loc}}$ is shifted downward a little.
\end{comment}
\label{Fig4}
\end{figure}
%===========================fig4===============================%
shows a schematic of the MOR, where the magnetic field $B$ is applied along with the propagation (i.e. $z$) direction, and the red (blue) solid circular curve with an arrow gives the polarization direction of the circular polarization component $\mathcal{E}_{p+}$ ($\mathcal{E}_{p-}$) of the probe field, which is orthogonal to the propagation direction.
The gray region is the atomic cell, and $\psi_{\rm tot}$ is the rotation angle due to the MOR after passing through the atomic cell.

Figure~\ref{Fig4}(b) shows MOR angles (in radian units) of $\psi_{\rm{lin}}$ (green dashed line), $\psi_{\rm{loc}}$ (yellow dotted line), $\psi_{\rm{nloc}}$ (blue dash-dotted line), and $\psi_{\rm{tot}}$ (red solid line) as functions of the magnetic field $B$ (in micro-tesla units) for $L=5$ ${\rm{mm}}$. When plotting the figure, the system parameters were taken to be
%%%%%%%%%%%%%%%%%%%%%%%
$\Delta_3=2\pi\times100$ $\mathrm{MHz}$, $\Delta_4=2\pi\times0.18$ $\mathrm{MHz}$, $\Omega_{{c}}=2\pi\times6.5$ $\mathrm{MHz}$,
$\mathcal{N}_a=3\times10^{10}$ $\mathrm{cm^{-3}}$,
and $A_1U_0=A_2U_0=2.74\times10^6$ ${\rm s}^{-1}$.

From the figure we see that $\psi_{\rm{lin}}$ and $\psi_{\rm{loc}}$ are a very small and have very weak dependence on the magnetic field $B$; however, $\psi_{\rm{nloc}}$ is strongly dependent on $B$ and increases rapidly as $B$ increases. In fact, $\psi_{\rm{nloc}}$  can take a sizable value and it is the main contributor to the total MOR angle $\psi_{\rm{tot}}$ when the magnetic field departs from zero.
For instance, for $B=0.9999\,{\rm\mu T}$ we obtain  the total MOR angle of the probe field
\begin{equation}
\psi_{\rm tot}=-0.2016\,{\rm rad}\,\,\,\, (\simeq -11.55\,^{\circ}),
\end{equation}
which is one order of magnitude larger than that obtained by using conventional EITs (without Rydberg-Rydberg interactions) where $\psi_{\rm tot}$ is estimated to be $0.0265\,{\rm rad}\,\, (\simeq 1.5192\,^{\circ}$) only.
Consequently, the Rydberg-Rydberg interaction in the system may be harnessed to realize a giant MOR of the probe field. Based on the giant MOR, one can design a magnetometer through double Rydberg EIT, by which very weak magnetic fields can be measured with very high precision.

%%%%%%%%%%%%%%%%%%%%%%%%%%%%%%%%%
\section{Summary}\label{sec5}

In this work we have investigated the Kerr nonlinearities and related magneto-optical effects for a probe laser field with two orthogonal polarization components, propagating in a cold Rydberg atomic gas with an inverted-Y-type level configuration and working under the condition of a double EIT. By means of an approach beyond both the MFA and the GSA, we have carried out systematic and detailed calculations on the third-order nonlinear optical susceptibilities and demonstrated that such a system supports giant nonlocal self- and cross-Kerr nonlinearities, which are contributed by the Rydberg-Rydberg interaction via the double EIT. The theoretical result of the cross-Kerr nonlinearity for $^{85}$Rb atomic gas is very close to the experimental one reported recently. In addition, we have shown that the probe laser field can gain a very large MOR under the action of an external weak magnetic field. The research results reported in the present work are useful for the development of nonlocal nonlinear magneto-optics and also have potential practical applications for precision measurements and for optical information processing and transmission, including the design of atomic magnetometers, quantum phase gates, few-photon bound states, quantum nondemolition measurement, nonlocal weak-light vector solitons and vortices, and so on.

%%%%%%%%%%%%%%%%%%%%%%%%%%%%%%%%%%%%%%%%%%%%%%%%%%%%%%%%%%%%%%
\acknowledgments

The authors thank Zhengyang Bai, Jiteng Sheng, and Jianming Zhao for fruitful discussions, and Josiah Sinclair for useful communications. This work was supported by the National Natural Science Foundation of China under Grant No. 11975098.

%%%%%%%%%%%%%%%%%%%%%%%%%%%%%%%%%%%%%%%%%%%%%%%%%%%%%%%%
\appendix

\section{Derivation of the Hamiltonian density}\label{appendix0}

For completeness, here we give a detailed derivation on the Hamiltonian density (\ref{eqn1}) in the main text.

\subsection{The Hamiltonian for a single atom interacting with the laser fields}

Under the electric dipole approximation, the Hamiltonian of a single atom at a given position ${\bf r}$ reads
\begin{eqnarray}
&& \hat{H}_{\rm single}=\hat{H}_0+\hat{H}', \,\,\,\,\hat{H}'=-{\bf p}\cdot {\bf E}({\bf r},t).
\end{eqnarray}
Here $\hat{H}_0$ is the Hamiltonian of the atom in the absence of the external optical field ${\bf E}$, $\hat{H}'$ is the interaction Hamiltonian between the atom and the laser field, {\bf p} is the electric dipole moment of the atom.
The laser field is $\mathbf{E}=\mathbf{E}_{p1}+\mathbf{E}_{p2}+\mathbf{E}_c$, with $\mathbf{E}_{pj}={\bf e}_{pj}{\cal E}_{pj}\exp [i(k_{pj}z-\omega_{pj} t)]+{\rm c.c.}$ ($j=1,2$) and $\mathbf{E}_c={\bf e}_c \mathcal{E}_{c}\exp [i(-k_{c}z-\omega_{c} t)]+\mathrm{c.c.}$.
Assuming $|\a\rangle$ is the eigen state of $\hat{H}_0$, i.e. $\hat{H}_0 |\a\rangle=E_{\a} |\a\rangle$\, ($\a=$1-4), we have
\begin{subequations}
\begin{eqnarray}
\hat{H}_{\rm single}
&& =\sum_\a E_\a |\a\rangle \langle \a|-{\bf E}({\bf r},t)\cdot
\sum_{\a,\b} {\bf p}_{\a\b} |\a\rangle \langle \b|\nonumber\\
&& =\sum_\a E_\a \hat{\sigma}_{\a\a}-{\bf E}({\bf r},t)\cdot
\sum_{\a,\b} {\bf p}_{\a\b} \hat{\sigma}_{\a\b},
\end{eqnarray}
\end{subequations}
where $\hat{\sigma}_{\a\b}=|\a\rangle \langle \b|$ is the transition matrix.
Introducing the transformation $\hat{\sigma}_{\a\b}({\bf r},t)=\hat{U}^{\dag}({\bf r},t)\hat{\sigma}_{\a\b}\hat{U}({\bf r},t)$, where
$\hat{U}({\bf r},t)$ is the evolution operator satisfying the Schr\"odinger equation $i\hbar\frac{\partial}{\partial t}\hat{U}({\bf r},t)=
\hat{H}_{\rm single}\hat{U}({\bf r},t)$, we obtain the Hamiltonian in the Heisenberg picture
\begin{eqnarray}
\hat{H}&& =\sum_\a E_\a \hat{\sigma}_{\a\a}({\bf r},t)\nonumber\\
&& -{\bf E}({\bf r},t)\cdot
\sum_{\a,\b} {\bf p}_{\a\b} \hat{\sigma}_{\a\b}({\bf r},t),\label{Hsingle}
\end{eqnarray}
with $\hat{\sigma}_{\a\b}({\bf r},t)$ satisfying the commutation relation
$[\hat{\sigma}_{\a\b}({\bf r},t),\hat{\sigma}_{\a'\b'}({\bf r}',t)]
=\d ({\bf r}-{\bf r}')[\delta_{\a'\b}\hat{\sigma}_{\a\b'}({\bf r},t)-\delta_{\a\b'}\hat{\sigma}_{\a'\b}({\bf r},t)]$.

%%%%%%%%%%%%%%%%%%%%%%%%%%%%%%%%%%%%%%%%
\subsection{The Hamiltonian and equation of motion for an atomic ensemble with Rydberg-Rydberg interaction}

Based on the above result, the Hamiltonian of an atomic ensemble with density ${\cal N}_a$ (assumed to be a constant for simplicity) reads
$\hat{H}_{\rm ensemble}=\sum_j\left[
\sum_\a E_\a \hat{\sigma}_{\a\a}({\bf r}_{j},t)-{\bf E}({\bf r}_{j},t)\cdot
\sum_{\a\b} {\bf p}_{\a\b} \hat{\sigma}_{\a\b}({\bf r}_{j},t)\right]$, which can be written in the form
%%%%%%%%%%%%%%%%%%%
\begin{eqnarray}
\hat{H}_{\rm ensemble}
&&={\cal N}_a \int d^3 r\left[
\sum_\a E_\a \hat{\sigma}_{\a\a}({\bf r},t)\right.\nonumber\\
&& \left. -{\bf E}({\bf r},t)\cdot
\sum_{\a,\b} {\bf p}_{\a\b} \hat{\sigma}_{\a\b}({\bf r},t)\right].
\end{eqnarray}

The Rydberg-Rydberg interaction energy between the atoms at the position ${\bf r}_j$ and ${\bf r}_{j'}$, respectively, is
$\hat{\s}_{44}({\bf r}_{j'},t)V_{\rm vdW}({\bf r}_{j'}-{\bf r}_{j})\hat{\s}_{44}({\bf r}_{j},t)$; the total energy due to the Rydberg-Rydberg interaction is
$\hat{H}_{\rm vdW}
=\sum_{j,j'}\hat{\s}_{44}({\bf r}_{j'},t)V_{\rm vdW}({\bf r}_{j'}-{\bf r}_{j})\hat{\s}_{44}({\bf r}_{j},t)=
{\cal N}_a^2 \int d^3 r \int d^3 r'
\hat{\s}_{44}({\bf r}',t)\hbar V({\bf r}'-{\bf r})\hat{\s}_{44}({\bf r},t)$,
where $\hbar V({\bf r}'-{\bf r})\equiv V_{\rm vdW}({\bf r}'-{\bf r})$ is the van der Waals interaction potential, with ${\bf r}'\neq {\bf r}$~\cite{Murray2016-1}. Hence, in the presence of the Rydberg-Rydberg interaction, the Hamiltonian of the atomic ensemble is given by $\hat{H}_{\rm total}= \hat{H}_{\rm essemble}+\hat{H}_{\rm vdW}={\cal N} \int d^3 r {\cal H}_{\rm total}({\bf r},t)$, where
\begin{eqnarray}\label{Htotal0}
{\cal H}_{\rm total}({\bf r},t)
&& =\sum_\a E_\a \hat{\sigma}_{\a\a}({\bf r},t)-{\bf E}({\bf r},t)\cdot
\sum_{\a\b} {\bf p}_{\a\b} \hat{\sigma}_{\a\b}({\bf r},t)\nonumber\\
&& +{\cal N}_a \int d^3 r'\hat{\s}_{44}({\bf r}',t)\hbar V({\bf r}'-{\bf r})\hat{\s}_{44}({\bf r},t)
\end{eqnarray}
is the Hamiltonian density. The Heisenberg equation of  motion for $\hat{\s}_{\a\b}({\bf r},t)$ reads
\begin{eqnarray}\label{HEM0}
i\hbar\frac{\partial}{\partial t}\hat{\sigma}_{\a\b}({\bf r},t)=
\left[\hat{\sigma}_{\a\b}({\bf r},t), \hat{H}_{\rm total} \right].
\end{eqnarray}

%%%%%%%%%%%%%%%%%%%%%
\begin{widetext}
Making the transformation $\hat{\sigma}_{\a\b}({\bf r},t)
=\hat{S}_{\b\a}({\bf r},t)\,e^{i\left[({\bf k}_\b-{\bf k}_\a)\cdot {\bf r}-
\left(\frac{E_\b-E_\a}{\hbar}+\Delta_\b-\Delta_\a\right)t \right]}$,
with ${\bf k}_{1}=0$, ${\bf k}_{2}={\bf k}_{p1}-{\bf k}_{p2}$,
${\bf k}_{3}={\bf k}_{p1}$, ${\bf k}_{4}={\bf k}_{p1}+{\bf k}_{c}$;
$\Delta_1=0$,
$\Delta_2=\omega_{p1}-(E_3-E_{1})/\hbar$, $\Delta_{2}=\omega_{p1}-\omega_{p2}-(E_{2}-E_{1})/\hbar$, and
$\Delta_4=\omega_c+\omega_{p1}-(E_4-E_{1})/\hbar$, the
Hamiltonian density becomes
%%%%%%%%%%%%%%%%%%%%%%%%%%%%%
%%%%%%%%%%%%%%%%%%%%%%%
\begin{eqnarray}\label{Htotal1}
{\cal H}_{\rm total}({\bf r},t)
=&& \sum_\a E_\a \hat{S}_{\a\a}({\bf r},t)
-\sum_{\a\b} ({\bf p}_{\a\b}\cdot {\bf E})\hat{S}_{\a\b}({\bf r},t)
e^{i\left[({\bf k}_\a-{\bf k}_\b)\cdot {\bf r}-
\left(\frac{E_\a-E_\b}{\hbar}+\Delta_\a-\Delta_\b\right)t \right]}\nonumber\\
&& +{\cal N}_a \int d^3 r'\hat{S}_{44}({\bf r}',t)\hbar V({\bf r}'-{\bf r})\hat{S}_{44}({\bf r},t),
\end{eqnarray}
which is the Hamiltonian density.
The Heisenberg equation (\ref{HEM0}) becomes
%The Heisenberg equation of  motion for $\hat{S}_{\a\b}({\bf r},t)$ is
\begin{eqnarray}\label{HEM1}
&& \left[i\frac{\partial}{\partial t}+\left(\frac{E_\a-E_\b}{\hbar}+\Delta_\a-\Delta_\b\right)\right]
\hat{S}_{\a\b}({\bf r},t)=\frac{1}{\hbar}
\left[\hat{S}_{\a\b}({\bf r},t), \hat{H}_{\rm total} \right].
\end{eqnarray}

%%%%%%%%%%%%%%%%%%%%%%%%%%%%%%%%%%%%%%%%%%%%
\subsection{The Hamiltonian under rotating-wave approximation}

Because the probe and control fields have a near-resonance interaction with the atoms, one can make a RWA to simply the Hamiltonian~\cite{Boyd2008}. By keeping only the near-resonance terms, Eq.~(\ref{Htotal1}) is reduced to the RWA Hamiltonian:
\begin{eqnarray}\label{HRWA10}
\hat{{\cal H}}_{\rm RWA} ({\bf r},t)\equiv && \sum_\a E_\a \hat{S}_{\a\a}({\bf r},t)
-\hbar\left[\Omega_{p1}^{*}\hat{S}_{31}({\bf r},t)+\Omega_{p2}^{*}\hat{S}_{32}({\bf r},t)
+\Omega_{c}^{*}\hat{S}_{43}({\bf r},t)+{\rm H.c.}\right]\nonumber\\
&& + {\cal N}_a \int d^3 r'\hat{S}_{44}({\bf r}',t)\hbar V({\bf r}'-{\bf r})\hat{S}_{44}({\bf r},t).
\end{eqnarray}

%%%%%%%%%%%%%%%%%%%%
For convenience, we define a Hamiltonian density
\begin{eqnarray}\label{HRWA}
\hat{{\cal H}} ({\bf r},t)
&&\equiv -\sum_{\a}(E_{\a}+\hbar\D_{\a})\hat{S}_{\a\a}({\bf r},t)+\hat{{\cal H}}_{\rm RWA}({\bf r},t)\nonumber\\
&&=-\sum_{\a}\hbar\D_{\a}\hat{S}_{\a\a}
-\hbar\left[\Omega_{p1}^{*}\hat{S}_{31}({\bf r},t)+\Omega_{p2}^{*}\hat{S}_{32}({\bf r},t)
+\Omega_{c}^{*}\hat{S}_{43}({\bf r},t)+{\rm h.c.}\right]\nonumber\\
&& \hspace{4mm}+ {\cal N}_a \int d^3 r'\hat{S}_{rr}({\bf r}',t)\hbar V({\bf r}'-{\bf r})\hat{S}_{rr}({\bf r},t),
\end{eqnarray}
which is the one given in Eq.~(\ref{eqn1}). With the total Hamiltonian  $\hat{H}={\cal N}_a \int d^3 r \hat{{\cal H}} ({\bf r},t)$,   Eq.~(\ref{HEM1}) is reduced to the simplified form
\begin{eqnarray}\label{HeiEq2}
&& i\frac{\partial}{\partial t}\hat{S}_{\a\b}({\bf r},t)
=\frac{1}{\hbar}\left[\hat{S}_{\a\b}, \hat{H}\right].
\end{eqnarray}

\end{widetext}

\begin{widetext}
%%%%%%%%%%%%%%%%%%%%%%%%%%%%%%%%
\section{Explicit expression of the optical Bloch equation}\label{appendixA}

Based on the Heisenberg equation of motion (\ref{HeiEq2}) and taking
${\bf e}_{p1}=\hat{\epsilon}_{+}$, ${\bf e}_{p2}=\hat{\epsilon}_{-}$, ${\bf e}_c=\hat{\epsilon}_c$, ${\cal E}_{p1}={\cal E}_{p+}$, ${\cal E}_{p2}={\cal E}_{p-}$, $k_{p1}=k_{p2}=k_p$,
and $\omega_{p1}=\omega_{p2}=\omega_p$, we can obtain an optical Bloch equation for the DM elements $\rho_{\alpha\beta}({\bf r},t)=\langle\hat {S}_{\alpha\beta}({\bf r},t)\rangle$  ($\alpha,\beta=$1-4), with the explicit form given by
%%%%%%%%%%%%%%%%%%%%%%%%%%%%%%%%%%%%%%%%%%%%%
\begin{subequations}
\begin{align}\label{eqn2}
&i\frac{\partial} {\partial t}\rho_{11}+i\Gamma_{21}\rho_{11}-i\Gamma_{12}\rho_{22}-i\Gamma_{13}
\rho_{33}+\Omega_{{p}1}^{\ast}\rho_{31}-\Omega_{{p}1}\rho_{13}=0,\\
%%%%%%%%%%%%%%%%%%%%%%%%%%%%%%%%%%%%%%%%%%
&i\frac{\partial} {\partial t}\rho_{22}+i\Gamma_{12}\rho_{22}-i\Gamma_{21}\rho_{11}-i\Gamma_{23}
\rho_{33}+\Omega_{{p}2}^{\ast}\rho_{32}-\Omega_{{p}2}\rho_{23}=0,\\
%%%%%%%%%%%%%%%%%%%%%%%%%%%%%%%%%%%%%%%%%
&i\frac {\partial} {\partial t}\rho_{33}+i\Gamma_3\rho_{33}-i\Gamma_{34}\rho_{44}-\Omega_{{p}1}^{\ast}\rho_{31}
+\Omega_{{p}1}\rho_{13}
-\Omega_{{p}2}^{\ast}\rho_{32}+\Omega_{{p}2}\rho_{23}
+\Omega_{{c}}^{\ast}\rho_{43}-\Omega_{{c}}\rho_{34}=0,\\
%%%%%%%%%%%%%%%%%%%%%%%%%%%%%%%%%%%%%%%
&i\frac {\partial} {\partial t}\rho_{44}+i\Gamma_{34}\rho_{44}-\Omega_{{c}}^{\ast}\rho_{43}
+\Omega_{{c}}\rho_{34}=0,
\end{align}
for  diagonal matrix elements, and
\begin{align}
%%%%%%%%%%%%%%%%%%%%%%%%%%%%%%%%%%%%%%%%%%%%%
&\left(i\frac {\partial} {\partial t}+d_{21}\right)\rho_{21}+\Omega_{{p}2}^{\ast}\rho_{31}-\Omega_{{p}1}
\rho_{23}=0,\\
%%%%%%%%%%%%%%%%%%%%%%%%%%%%%%%%%%%%%%%%%%%
&\left(i\frac {\partial} {\partial t}+d_{31}\right)\rho_{31}+\Omega_{{c}}^{\ast}\rho_{41}
+\Omega_{{p}1}\left(\rho_{11}-\rho_{33}\right)+\Omega_{{p}2}
\rho_{21}=0,\\
%%%%%%%%%%%%%%%%%%%%%%%%%%%%%%%%%%%%%%%%%%
&\left(i\frac {\partial} {\partial t}+d_{32}\right)\rho_{32}+\Omega_{{c}}^{\ast}\rho_{42}
+\Omega_{{p}2}\left(\rho_{22}-\rho_{33}\right)
+\Omega_{{p}1}\rho_{12}=0,\\
%%%%%%%%%%%%%%%%%%%%%%%%%%%%%%%%%%%%%%%%%
&\left(i\frac {\partial }{\partial t}+d_{41}\right)\rho_{41}+\Omega_{{c}}\rho_{31}-\Omega_{{p}1}\rho _{43}-\mathcal{N}_a\int d^3r' V\left(\mathbf {r}'-\mathbf {r}\right)\rho_{44,41}\left( \mathbf {r'},\mathbf {r},t \right)=0,\\
%%%%%%%%%%%%%%%%%%%%%%%%%%%%%%%%%%%%%
&\left(i\frac{\partial}{\partial t }+d_{42}\right)\rho_{42}+\Omega_{{c}}\rho_{32}-\Omega_{{p}2}
\rho_{43}-\mathcal {N}_{a }\int d^3r'V\left(\mathbf{r}'-\mathbf{r}\right)
\rho_{44,42}\left(\mathbf{r}',\mathbf{r},t\right)=0,\\
%%%%%%%%%%%%%%%%%%%%%%%%%%%%%%%%%%%%%%%%
&\left(i\frac{\partial}{\partial t}+d_{43}\right)\rho_{43}+\Omega_{{c}}\left(\rho_{33}-\rho_{44}\right)
-\Omega_{{p}1}^{\ast }\rho_{41}-\Omega_{{p}2}^{\ast}\rho_{42}
-\mathcal{N}_{a}\int  d^3r' V\left(\mathbf{r}' -\mathbf{r}\right)\rho_{44,43}\left(\mathbf{r}',\mathbf{r},t\right)=0,
\end{align}
\end{subequations}
for nondiagonal matrix elements. Here
%%%%%%%%%%%%%%%%%%%%%%%%%%%%%%%%%%%%%%%%%%%%%
$d_{\alpha\beta }=\Delta _{\alpha }-\Delta _{\beta}+i\gamma_{\alpha\beta }$, $\gamma _{\alpha \beta}=\left(\Gamma _{\alpha }+\Gamma_{\beta} \right)/2+\gamma _{\alpha\beta}^{\rm{dep}}$,
%%%%%%%%%%%%%%%%%%%%%%%
$\Gamma_{\beta}=\sum_{\alpha <\beta}\Gamma _{\alpha\beta}$
%%%%%%%%%%%%%%%%%%%%%%%%
$\Gamma_{\alpha\beta}$ is the spontaneous emission decay rate from the state $|\beta\rangle$ to the state $|\alpha\rangle$, $\gamma_{\alpha\beta}^{\rm{dep}}$ is the dephasing rate reflecting the loss of phase coherence between
%%%%%%%%%%%%%%%%%%%%%%%%
$|\alpha\rangle$ and $|\beta\rangle$, and
$V_{\mathrm{vdW}} \equiv \hbar V\left(\mathbf{r}'-\mathbf{r}\right)$
%%%%%%%%%%%%%%%%%%%%%%%
is the van der Waals interaction potential between two Rydberg atoms
located at position ${\bf r}$ and ${\bf r}'$. Note that the above equations for one-body DM elements involve two-body DM elements
%%%%%%%%%%%%%%%%%%%%%%%
$\rho_{44,4\alpha} \left(\mathbf{r}',\mathbf{r}, t\right)=\langle\hat{S}_{44}\left(\mathbf {r}', t\right)\hat{S}_{4\alpha}( \mathbf{r},t)\rangle$~($\alpha=1,2,3$).
\end{widetext}

%%%%%%%%%%%%%%%%%%%%%%%%%%%%%%%%%%%%%%%%%%
\section{Steady-state solutions of the Bloch equation up to third-order approximation}\label{appendixB}

The expressions of the Kerr nonlinear susceptibilities of the system can be obtained by solve the Bloch equation (\ref{eqn2}) under a steady-state approximation (i.e., taking $\partial/\partial t=0$).
%To calculate the Kerr nonlinear susceptibilities of the system, solution of %the Bloch equation is needed up to third-order approximation.
To this end, we make the perturbation expansion
%%%%%%%%%%%
$\rho_{\alpha\alpha}=\rho_{\alpha\alpha}^
{(0)}+\epsilon\rho_{\alpha\alpha}^{(1)} +\epsilon^{2}\rho_{\alpha\alpha}^{(2)} +\cdots$ ($\alpha=1,2,3,4$), and
%%%%%%%%%%%%%%%%%%%%%%%%%
$\rho_{\alpha\beta} =\epsilon\rho_{\alpha\beta}^{(1)}+\epsilon^{2}\rho _{\alpha\beta}^{(2)}+\cdots$ ($\alpha= 2,3,4$; $\beta=1,2,3$; and $\beta<\alpha$),
%%%%%%%%%%%%%%%%%%%%%%%%%%%%%
where $\epsilon$ is a dimensionless small parameter characterizing the typical amplitude of the probe field, i.e.,
$\Omega_{p1}\approx \Omega_{p2}\sim \epsilon$.

%%%%%%%%%%%%%%%%%%%%%%%%%%%%%%%%%%
\subsection{Zeroth- and first-order solutions}\label{appendixB2}

Substituting the perturbation expansion described above into the Bloch equation (\ref{eqn2}), we get the zeroth-order solution $\rho_{11}^{(0)}+\rho_{22}^{(0)}=1$  with all the other $\rho_{jl}^{(0)}$ equal to zero. Here we assume $\rho_{11}^{(0)}=\rho_{22}^{(0)}=1/2$, i.e., the initial population of atoms is prepared in the two ground states $|1\rangle$ and $|2\rangle$.

At the first order of the expansion, the solution for $\rho_{31}^{(1)}$, $\rho_{41}^{(1)}$, $\rho_{32}^{(1)}$, and $\rho_{42}^{(1)}$ reads
\begin{equation}\label{rho1}
\rho_{ij}^{(1)}={a}_{ij}^{(1)}\Omega_{{pj}},
\end{equation}
where ${a}_{31}^{(1)}=d_{41}/(2D_1)$, ${a}_{32}^{(1)}=d_{42}/(2D_2)$, ${a}_{41}^{(1)}=-\Omega_c/(2D_1)$, ${a}_{42}^{(1)}=-\Omega_c/(2D_2)$, and $D_j=|\Omega_{{c}}|^2-d_{4j}d_{3j}$.

\subsection{Second-order solution}\label{appendixB3}

At the second order of the expansion, nonzero matrix elements $ \rho_{11}^{(2)}$, $\rho_{22}^{(2)}$, $\rho_{33}^{(2)}$, $\rho_{44}^{(2)}$, and $ \rho_{43}^{(2)}$ satisfy the equations
\begin{align}\label{rho211}
&\left( \begin{matrix}
i\Gamma_{21} &-i\Gamma_{12}&-i\Gamma_{13}&0&0&0\\
0&0& i\Gamma_3& -i\Gamma_{34}& \Omega_c^*&-\Omega_c\\
0&0&0& i\Gamma_{34}&-\Omega_c^*&\Omega_c\\
1& 1& 1& 1& 0&0\\
0&0&\Omega_c&-\Omega_c&d_{43}&0\\
0&0&\Omega_c^*&-\Omega_c^*&0&d_{43}^*\end{matrix} \right)
\left( \begin{matrix} \rho_{11}^{(2)}\\ \rho_{22}^{(2)}\\\rho_{33}^{(2)}\\\rho_{44}^{(2)}\\  \rho_{43}^{(2)}\\\rho_{34}^{(2)}\end{matrix} \right)
\notag\\
&
=\left( \begin{matrix}
\Omega_{{p}1}\rho_{13}^{(1)}-\Omega _{{p}1}^{\ast}\rho_{31}^{(1)}\\
%\Omega_{{p}1}^*\rho_{31}^{(1)}-\Omega_{{p}1}\rho_{13}^{(1)}+\Omega_{{p2}}^*\rho_{32}^{(1)}-\Omega_{{p}2}\rho_{23}^{(1)}\\
i\Omega_{{p}1}^*\rho_{31}^{(1)}-i\Omega_{{p}1}\rho_{13}^{(1)}+\Omega_{{p2}}^*\rho_{32}^{(1)}-\Omega_{{p2}}\rho_{23}^{(1)}\\
0\\
0\\
\Omega_{p1}^*\rho_{41}^{(1)}+\Omega_{p2}^*\rho_{42}^{(1)}\\
\Omega_{p1}\rho_{14}^{(1)}+\Omega_{p2}\rho_{24}^{(1)}
\end{matrix} \right),
\end{align}
Their solution has the form $\rho_{\alpha\alpha}^{(2)}=a_{\alpha\alpha,1}^{(2)}|\Omega_{{p1}}|^{2}
+a_{\alpha\alpha,2}^{(2)}|\Omega_{{p2}}|^{2}$, with the coefficients given by
%%%%%%%%%%%%%%%%%%%%%%%%%%%
\begin{subequations}\label{eqs1}
\begin{align}
&a_{33,j}^{(2)}=\frac{A_{j}}{i\Gamma_{3}},\quad
\\&
a_{44,j}^{(2)}=\frac{|\Omega_c|^2(A_{j}^*d_{43}-d_{43}^*B_{j})-|d_{43}|^2A_{j}
+C_ja_{33,j}^{(2)}}{i[2\gamma_{43}|\Omega_c|^2+|d_{43}|^2\Gamma_{34}]},
\\&
a_{11,j}^{(2)}=\frac{-A_{j}+i(\Gamma_{13}-\Gamma_{12})a_{33,j}^{(2)}
-i\Gamma_{12}a_{44,j}^{(2)}}{i(\Gamma_{12}+\Gamma_{21})},
\\&
a_{22,j}^{(2)}=\frac{-A_{j}+i(\Gamma_{13}+\Gamma_{21})a_{33,j}^{(2)}
+i\Gamma_{21}a_{44,j}^{(2)}}{-i(\Gamma_{12}+\Gamma_{21})},
\\&
a_{43,j}^{(2)}=\frac{-{\Omega_c}B_{j}-a_{33,j}^{(2)}+a_{44,j}^{(2)}}{d_{43}},
\end{align}
\end{subequations}
where $A_{j}=a_{3j}^{(1)}-a_{3j}^{(1)*},$ $B_{j}=1/({2D_{j}})$, $C_j=i\Gamma_{3}|d_{43}|^2+2i\gamma_{43}|\Omega_c|^2$.
%%%%%%%%%%%%%%%%%%%%%%%%%%%%%%

The expression of $\rho_{21}^{(2)}$ reads
\begin{align}\label{12}
\rho_{21}^{(2)}&=\frac{\Omega_{{p}1}\rho_{23}^{(1)}-\Omega_{{p}2}^{*}\rho_{31}^{(1)}}{\omega+d_{21}}\equiv a_{21}^{(2)}\Omega_{p1}\Omega_{p2}^*.
\end{align}

\subsection{Third-order solution}\label{appendixB4}

The solutions of $\rho_{3j}^{(3)}$ and $\rho_{4j}^{(3)}$ are obtained by solving the equations
\begin{align}
&\left(\begin{matrix}
 d_{3j}&\Omega_{c}^{\ast}\\
\Omega_c&d_{4j}\end{matrix}\right)
\left(\begin{matrix}
\rho_{3j}^{(3)}\\
\rho_{4j}^{(3)} \\
\end{matrix}\right)\notag\\
&=\left(\begin{matrix}
{-\Omega_{pj}\left(\rho_{jj}^{(2)}-\rho_{33}^{(2)}\right)-\Omega_{p,3-j}\rho_{21}^{(2)}} \\
{\Omega_{pj} \rho_{43}^{(2)}+\mathcal{N}_{a} \int d^3r' \mathrm{V}\left(\bf{r}^{\prime}-\bf{r}\right) \rho_{44,4j}^{(3)}\left(\bf{r}^{\prime}- \bf{r}\right) }\end{matrix}\right),
\end{align}
with $\rho_{44,4l}^{(3)}\left(\bf{r}^{\prime}- \bf{r}\right)=a_{44,4l,1}^{(3)}\left({\bf r}^{\prime}, {\bf r}, t\right)|\Omega_{{p1}}({\bf r}^{\prime})|^{2}\Omega_{{pl}}({\bf r}^{\prime})
+a_{44,4l,2}^{(3)}\left({\bf r}^{\prime}, {\bf r}, t\right)|\Omega_{{p2}}({\bf r}^{\prime})|^{2}\Omega_{{pl}}(\bf{r})$. Then we have
\begin{align}\label{rho3}
\rho_{jl}^{(3)}&=a_{jl,1}^{(3,\rm{loc})}|\Omega_{{p1}}|^{2}\Omega_{{pl}}
+a_{jl,2}^{(3,\rm{loc})}|\Omega_{{p2}}|^{2}\Omega_{{pl}}\notag\\
&+\int d^3r' a_{jl,1}^{(3,\rm{nloc})}|\Omega_{{p1}}({\bf r}^{\prime})|^{2}\Omega_{{pl}}({\bf r})\notag\\
&+\int d^3r' a_{jl,2}^{(3,\rm{nloc})}|\Omega_{{p2}}({\bf r}^{\prime})|^{2}\Omega_{{pl}}({\bf r}),
\end{align}
with the coefficients given by
\begin{subequations}\label{rho41local}
\begin{align}
&a_{31,1}^{(3,\rm{loc})}=\frac{\Omega_{c}^{\ast}a_{43,1}^{(2)}+d_{41}\left(a_{11,1}^{(2)}-a_{33,1}^{^{(2)}}\right)}{D_1},\\
&a_{31,2}^{(3,\rm{loc})}=\frac{\Omega_{c}^{\ast}a_{43,2}^{(2)}+d_{41}\left(a_{11,2}^{(2)}-a_{33,2}^{^{(2)}}+X\right)}{D_1},\label{411}\\
&a_{32,2}^{(3,\rm{loc})}=\frac{\Omega_{c}^{\ast}a_{43,2}^{(2)}+d_{42}\left(a_{22,2}^{(2)}-a_{33,2}^{^{(2)}}\right)}{D_2},\\ &a_{32,1}^{(3,\rm{loc})}=\frac{\Omega_{c}^{\ast}a_{43,1}^{(2)}+d_{42}\left(a_{22,1}^{(2)}-a_{33,1}^{^{(2)}}+X^*\right)}{D_2},\label{422}\\
%%%%%%%%%%%%%%%%%%%%
&a_{3j,l}^{(3,\rm{nloc})}=\frac{\Omega_{{c}}^{\ast}\mathcal{N}_{a} \mathrm{V}\left({\bf r}^{\prime}-{\bf r}\right) a_{44,4j,l}^{(3)}\left({\bf r}^{\prime}- {\bf r}\right)}{D_j},
\end{align}
\end{subequations}
where $X=(a_{23}^{(1)}-a_{31}^{(1)})/d_{21}$. Note that the solution $\rho_{31}^{(3)}$ and $\rho_{32}^{(3)}$ given by (\ref{rho3}) includes the parts of local terms and nonlocal (integral) terms contributed by the Rydberg-Rydberg interaction; to get the nonlocal terms we must solve the equations of motion for two-body DM elements $\rho_{\alpha\beta,\mu\nu}$, which are yet to be determined.

\section{Steady-state solutions of the equations for two-body DM elements}\label{appendixB5}

\subsection{Second-order solution}
The nonzero solution of two-body DM elements starts at the second-order approximation. The two-body DM elements $\rho_{41,41}^{(2)}$ and $\rho_{42,42}^{(2)}$
satisfy the equation
%%%%%%%%%%%%%%%%%%%%%
{\small \begin{align}\label{rho4141}
&&\left(\begin{matrix}{2 d_{4\alpha}-V} & {2 \Omega_{{c}}} & {0} \\ {\Omega_{{c}}^{*}} & {d_{3\alpha}+d_{4\alpha}}  & {\Omega_{{c}}}  \\ {0} & {2\Omega_{{c}}^{*}} & {2d_{3\alpha}}\end{matrix}\right)
\left(\begin{matrix}{\rho_{4\alpha,4\alpha}^{(2)}} \\ {\rho_{4\alpha,3\alpha}^{(2)}} \\ {\rho_{3\alpha,3\alpha}^{(2)}} \end{matrix}\right)
 =\left(\begin{matrix}{0} \\ {-\Omega_{p\alpha}\rho_{4\alpha}^{(1)}/2} \\ -\Omega_{p\alpha}{\rho_{3\alpha}^{(1)}}\end{matrix}\right),
\end{align}}\noindent
where $\alpha=1,\,2$, and $\rho_{42,41}^{(2)}$ satisfies the equation
\begin{align}
&\left(\begin{matrix}
M & {\Omega_{{c}}} & {\Omega_{{c}}} & {0} \\
{\Omega_{{c}}^{*}} & {d_{42}+d_{31}} & {0} & {\Omega_{{c}}}  \\
{\Omega_{{c}}^{*}} & {0}  & {d_{41}+d_{32}} & {\Omega_{{c}}}  \\
{0} & {\Omega_{c}} & {\Omega_{{c}}^{*}} &  {d_{32}+d_{31}} \end{matrix}\right)
\left(\begin{matrix}{\rho_{42,41}^{(2)}} \\ {\rho_{42,31}^{(2)}} \\ {\rho_{41,32}^{(2)}} \\ {\rho_{32,31}^{(2)}}\end{matrix}\right)\notag\\
&=\left(\begin{matrix}
{0} \\ {-\Omega_{{p1}}\rho_{42}^{(1)}/2}\\
{-\Omega_{{p2}}\rho_{41}^{(1)}/2} \\ {-\Omega_{{p1}}\rho_{32}^{(1)}/2-\Omega_{{p2}}\rho_{31}^{(1)}/2} \end{matrix}\right),
\end{align}
with $M=d_{42}+d_{41}-V\left({\bf r}^{\prime}- {\bf r}\right)$.

\subsection{Third-order solution}

The third-order two-body DM elements can be obtained by solving the equations ($\alpha=1,\,2$)
 \begin{widetext}
%%%%%%%%%%%%%%%%%%%%%
{\small \begin{align}\label{third-order two-body}
&\left(\begin{matrix}
{d_{3\alpha}+i\Gamma_3} & {\Omega_{c}^*} & {-i\Gamma_{34}}&{\Omega_{c}^*} &{0}&{0}&{-\Omega_{c}}&{0}  \\
%%%%%%%%%%%%%%
{\Omega_{{c}}} & {d_{3\alpha}+d_{43}}  & {-\Omega_{c}}& {0}&{\Omega_{c}^*}& {0}& {0}& {0}  \\
%%%%%%%%%%%%%%%%%%
{0} & {-\Omega_{{c}}^{*}} & {d_{3\alpha}+i\Gamma_{34}}&{0}&{0}&{\Omega_{c}^*}&{\Omega_{c}}&{0}\\
%%%%%%%%%%%%%%%%%%%
{\Omega_{c}}& {0}& {0}&{i\Gamma_{3}+d_{4\alpha}}&{\Omega_{c}^*}&{-i\Gamma_{34}}& {0}& {-\Omega_{c}}\\
%%%%%%%%%%%%%%%%%
{0}&{\Omega_{c}}& {0}&{\Omega_{c}}&{d_{4\alpha}+d_{43}}-V&{-\Omega_{c}}& {0}& {0}\\
%%%%%%%%%%%%%%%%%%
{0}& {0}&{\Omega_{c}}& {0}&{-\Omega_{c}^*}&{i\Gamma_{34}+d_{4\alpha}}-V&{0}&{\Omega_{c}}\\
%%%%%%%%%%%%%%%%%
{\Omega_{{c}}^{*}} &{0}&{-\Omega_{{c}}^{*}}&{0}&{0}&{0}&{d_{3\alpha}-d_{43}^*}&{\Omega_{c}^*}\\
%%%%%%%%%%%%%%%%%
{0}&{0}&{0}&{-\Omega_{{c}}^{*}} &{0}&{\Omega_{{c}}^{*}}&{\Omega_{{c}}}&{-d_{43}^*+d_{4\alpha}}
\end{matrix}\right)
%%%%%%%%%%%%%%%%%%%%
\left(\begin{matrix}
{\rho_{3\alpha,33}^{(3)}} \\ {\rho_{3\alpha,43}^{(3)}} \\ {\rho_{3\alpha,44}^{(3)}} \\
{\rho_{4\alpha,33}^{(3)}} \\ {\rho_{4\alpha,43}^{(3)}} \\ {\rho_{4\alpha,44}^{(3)}} \\
{\rho_{3\alpha,34}^{(3)}} \\ {\rho_{4\alpha,34}^{(3)}} \\
\end{matrix}\right)
 \nonumber\\
& =\left(\begin{matrix}
{\Omega_{p1}^*\rho\rho_{31,3\alpha}^{(2)}
-\Omega_{p1}d_{41}^*\rho_{3\alpha}^{(1)}/D_1^*
+\Omega_{p2}^*\rho\rho_{32,3\alpha}^{(2)}
-\Omega_{p2}d_{42}^*\rho_{3\alpha}^{(1)}/D_2^*-\Omega_{p1}\rho_{33}^{(2)}/2} \\ {\Omega_{p1}^*\rho\rho_{41,3\alpha}^{(2)}+\Omega_{p2}^*\rho\rho_{42,3\alpha}
-\Omega_{p1}\rho_{43}^{(2)}/2} \\
{-\Omega_{p1}\rho_{44}^{(2)}/2}
\\
{\Omega_{p1}^*\rho\rho_{31,4\alpha}^{(2)}-\Omega_{p1}d_{41}^*\rho_{4\alpha}/D_1^*
+\Omega_{p2}^*\rho\rho_{32,4\alpha}-\Omega_{p2}^*d_{42}^*\rho_{4\alpha}^{(1)}/D_2^*}
\\
{\Omega_{p1}^*\rho\rho_{41,4\alpha}^{(2)}+\Omega_{p2}^*\rho\rho_{42,4\alpha}^{(2)}}
\\
{0}
\\
{-\Omega_{p1}\rho_{14}^{(1)}\rho_{3\alpha}^{(1)}-\Omega_{p2}
\rho_{24}^{(1)}\rho_{3\alpha}^{(1)}-\Omega_{p1}\rho_{34}^{(2)}/2}
\\
{-\Omega_{p1}\rho_{14}^{(1)}\rho_{4\alpha}^{(1)}-\Omega_{p2}\rho_{24}^{(1)}
\rho_{4\alpha}^{(1)}}
\end{matrix}\right).
\end{align}
Through solving them, one can obtain the general expression of $a_{44,4j,l}^{(3)}\left(\bf{r}^{\prime}-\bf{r}\right)$, which reads
%%%%%%%%%%%%%%%%%%%%%%%%%%%%%%%%%%%%%%
\begin{align}\label{eqn12}
&a_{44,4j,l}^{(3)}\left(\bf{r}^{\prime}-\bf{r}\right)=\frac{P_{0jl}+P_{1jl}V\left( \bf{r}^{\prime}-\bf{r}\right)+P_{2jl}V\left( \bf{r}^{\prime}-\bf{r}\right)^2}{Q_{0jl}\!+\!Q_{1jl}V\left( \bf{r}^{\prime}-\bf{r}\right)\!+\!Q_{2jl}V\left( \bf{r}^{\prime}-\bf{r}\right)^2\!+\!Q_{3jl}V\left( \bf{r}^{\prime}-\bf{r}\right)^3},
\end{align}}\noindent
%%%%%%%%%%%%%%%%%%%%%%%%%%%%%%%%%%%%%%%
where $P_{ajl} \text{ and }Q_{ajl}\,(a=0,1,\,2,\,3)$ are functions of $\Gamma_{\alpha\beta}$, $\Delta_{\alpha}$, and $\Omega_{{c}}$. Their explicit expressions are lengthy and thus are omitted here.
\end{widetext}

%%%%%%%%%%%%%%%%%%%%%%%%%%%%%%%
The electric polarization intensity of the probe field in the atomic gas is given by
${\bf P}_p=
{\cal N}_a \{({\bf p}_{13}\rho_{31}+{\bf p}_{23}\rho_{32})\exp [i({ k}_{p} z-\omega_{p} t)]+{\rm c.c.}\}$,
which can be expressed by
$\mathbf{P}_p=\varepsilon_{0} ({\bf{\hat{\epsilon}_{-}}} \mathcal{E}_{p-}\chi_{1}+{\bf{\hat{\epsilon}_{+}}}\mathcal{E}_{p+}\chi_{2})\exp [i(k_{p} z-\omega_{p} t)]+{\rm{c.c.}}$,
where
%%%%%%%%%%%%%%%%%
\begin{align}
\chi_{1(2)}=\frac{\mathcal{N}_{a}\left({\bf{\hat{\epsilon}_{-(+)}}}
\cdot\mathbf{p}_{13(23)}\right)\rho_{31(32)}}{\varepsilon_{0} \mathcal{E}_{p-(+)}},
\end{align}
are optical susceptibilities.

Based on the above results obtained by the perturbation expansions, we have
$\rho_{3j}\simeq a_{3j}^{(1)} \Omega_{{pj}}+\sum_{l}^{1,2}a_{3j,l}^{(3,\rm{loc})}|\Omega_{{pl}}|^{2} \Omega_{{pj}}+\int d^3r'\sum_{l}^{1,2}a_{3j,l}^{(3,\rm{nloc})}|\Omega_{{pl}}({\bf {r}}^{\prime})|^{2} \Omega_{{pj}}$ ($j=1,\,2$), with $\Omega_{p1(2)}=\left( {\bf \bm{p}}_{13(23)}\cdot\hat{\epsilon}_{-({+}}\right){\cal
E}_{p-(+)}/\hbar$. Then we obtain the explicit expressions of the optical susceptibilities
%%%%%%%%%%%%%%%%%%%%%%%%
\begin{eqnarray}\label{eqn10}
\chi_{j}=
& \chi_{j}^{(1)}+\chi_{j1}^{(3,\rm{loc})}\left|\mathcal{E}_{p-}\right|^{2}
+\chi_{j2}^{(3,\rm{loc})}
\left|\mathcal{E}_{{p+}}\right|^{2}\nonumber\\
&+\chi_{j1}^{(3,\rm{nloc})}\left|\mathcal{E}_{p-}\right|^{2}
+\chi_{j2}^{(3,\rm{nloc})}\left|\mathcal{E}_{{p+}}\right|^{2},
\end{eqnarray}
%%%%%%%%%%%%%%%%%%%%%%%%%%%
where $\chi_{j}^{(1)}={\mathcal{N}_{a}\left|\mathbf{p}_{j3}
\right|^{2}}a_{3j}^{(1)}/({\varepsilon_{0} \hbar})$ are the first-order (linear) susceptibilities, and $\chi_{jl}^{(3,\rm{loc})}={\mathcal{N}_{a}\left|\mathbf{p}_{j3}
\right|^{4}}a_{3j,l}^{(3,\rm{loc})}/({\varepsilon_{0} \hbar^3})$ and
$\chi_{jl}^{(3,\rm{nloc})}={\mathcal{N}_{a}\left|\mathbf{p}_{j3}
\right|^{4}}a_{3j,l}^{(3,\rm{nloc})}/({\varepsilon_{0} \hbar^3})$\,
($j,l=1,2$) are local and nonlocal third-order nonlinear susceptibilities, respectively.

\section{The derivation of the nonlinear envelope equations}\label{appd}

The envelope equations governing the nonlinear evolution of the two polarization components of the probe field can be obtained by means of the method of multiple-scales~\cite{Newell1990} to solve the Bloch equation.~(\ref{appendixA}) and the Maxwell equation.~(\ref{eqn3}), which are coupled together. To get these equations, we take the perturbation expansion to be the same as that described in Appendix~\ref{appendixB}, but also with the expansion for the half Rabi frequencies of the probe field, i.e., $\Omega_{pj} =\epsilon\Omega_{pj}^{(1)}+ \epsilon^{2}\Omega_{pj}^{(2)}+ \cdots$ ($j=1,2$). In order to consider the spatial-temporal evolution of the system, we assume that $\Omega_{pj}^{(m)}$ and $\rho_{\alpha\beta}^
{(m)}$ are functions of the multiple-scale variables $z_l=\epsilon^l z$ and $t_l=\epsilon^l t$
($l=0,2$).

Carrying out the calculation up to the third-order approximation and returning to the original variables, we obtain %the equations for the envelopes (i.e. $F_1$ and $F_2$) of the two components of the probe field
%%%%%%%%%%%%%%%%%%%%%%%%%%%%%%%%%%%%%%%%%%%%
\begin{subequations}\label{eqn5}
\begin{align}
&i\left(\frac{\partial F_{1}}{\partial z}+\frac{1}{V_{\mathrm{g}1}} \frac{\partial F_{1}}{\partial t}\right)-\left(W_{ 11}\left|F_{1}\right|^{2}+W_{12}\left|F_{2}\right|^{ 2}\right)F_{1}\notag\\
&-\int d^3r'\sum_{l=1}^2G_{1l}({\bf r}'-{\bf r})\left| F_{l}\left(\bf{r}^{\prime}\right) \right|^{2}
F_{1}\left(\bf{r}\right)= 0, \label{eqn51}\\
%%%%%%%%%%%%%%%%%%%%%%%%%%%%%%%%%%%%%%%%%%%
&i\left(\frac{\partial F_{2}} {\partial z}+\frac{1}{V_{\mathrm{g}2}} \frac{\partial F_{2}}{\partial t}\right)-\left(W_{21}\left|F_{1}\right|^{2}+W_{22} \left|F_{2}\right|^{2}\!\right) F_{2}\notag\\
&-\int d^3r' \sum_{l=1}^2G_{2l}({\bf r}'-{\bf r})\left|F_{l}\left(\bf{r}^{\prime}\right)
\right|^{2} F_{2}\left(\bf{r}\right)= 0, \label{eqn52}
\end{align}
\end{subequations}
%%%%%%%%%%%%%%%%%%%%%%%%%%%%%%%%%%%%%%%%%%%
where $\theta_{j}=K_j(\omega)z-\omega t$, $F_j=\Omega_{pj}^{(1)}\,e^{-i\theta_j}$ is the envelope of the $j$th  ($j=1,2$) polarization components of the probe field,
$V_{gj}=(\partial K_j/\partial \omega)^{-1}$ is the group velocity of the $j$th polarization component; $W_{jl}$ (proportional to $\chi_{jl}^{(3,\rm{loc})}$) are coefficients of local Kerr nonlinearities characterizing the SPMs for $j=l$ and CPMs for $j\neq l$, and $G_{jl}$ (proportional to $\chi_{jl}^{(3,\rm{nloc})}$) are coefficients of nonlocal Kerr nonlinearities characterizing the nonlocal SPMs for $j=l$ and CPMs for $j\neq l$.
Explicit expressions of $W_{jl}$ and $G_{jl}$ are given by
\begin{subequations}\label{W and G}
\begin{align}
&W_{11}=-\frac{\kappa_{14}}{D_1}\left[\Omega_{c}^{\ast}a_{43,1}^{(2)}+d_{41}\left(a_{11,1}^{(2)}-a_{33,1}^{^{(2)}}\right)\right],\\ &W_{12}=-\frac{\kappa_{14}}{D_1}\left[\Omega_{c}^{\ast}a_{43,2}^{(2)}+d_{41}\left(a_{11,2}^{(2)}-a_{33,2}^{^{(2)}}+X\right)\right],\\
&W_{22}=-\frac{\kappa_{24}}{D_2}\left[\Omega_{c}^{\ast}a_{43,2}^{(2)}+d_{42}\left(a_{22,2}^{(2)}-a_{33,2}^{^{(2)}}\right)\right],\\ &W_{21}=-\frac{\kappa_{24}}{D_2}\left[\Omega_{c}^{\ast}a_{43,1}^{(2)}+d_{42}\left(a_{22,1}^{(2)}-a_{33,1}^{^{(2)}}+X^*\right)\right],\\
%%%%%%%%%%%%%%%%%%%%
&G_{jl}=-\frac{\kappa_{j4}}{D_1}\left[\Omega_{{c}}^{\ast}\mathcal{N}_{a} \mathrm{V}\left({\bf r}^{\prime}- {\bf r}\right) a_{44,4j,l}^{(3)}\left({\bf r}^{\prime}- {\bf r}\right)\right],
\end{align}
\end{subequations}

Equations (\ref{eqn51}) and (\ref{eqn52}) can be written into the dimensionless form
%%%%%%%%%%%%%%%%%%%%%%%%%%%%%%%%%%%%%%%%%%%%
\begin{align}\label{eqn7}
i\frac{\partial{u_{j}}} {\partial s }&+(-1)^{j-1}ig_{\delta}\frac{\partial u_{j}}{\partial\sigma}-\left(\sum_{l=1}^2 w_{jl}\left|u_{l}\right|^{2}\right)u_{j}\notag\\
&-\int d^3\zeta'\sum_{l=1}^2g_{jl}(\vec{\zeta}'-\vec{\zeta})|u_{l}(\vec{\zeta}^{\prime})| ^{2}
u_{j}(\vec{\zeta})= 0.
\end{align}
%%%%%%%%%%%%%%%%%%%%%%%%%%%%%
Here we have defined $u_{j}=F_j/U_0$,  $s=z/L_{\rm NL}$, $\sigma=(t-z/V_g)/\tau_0$,  $\vec{\zeta}={\bf r}/R_0=(\xi,\eta,s)$, $V_g=2V_{g1}V_{g2}/(V_{g1}+V_{g2})$,
$g_{\delta}={\rm sign}(\delta)L_{\rm NL}/L_{\delta}$, $w_{jl}=W_{jl}/|W_{11}|$, and $ g_{jl}=G_{jl}/|W_{11}|$.
In these definitions, $U_0$ is the typical Rabi frequency, $\tau_0$  ($R_0$) is the typical temporal duration (transverse size) of the probe field; $L_{\rm NL}=1/(|W_{11}|U_0^2)$ is the nonlinearity length; $L_{\delta}=\tau_0/|\delta|$ is the length of group-velocity mismatch, and  $\delta=(1/V_{g1}-1/V_{g2})/2$ is the parameter characterizing the group-velocity mismatch.

Assuming $\tau_0=9\times10^{-6}\,{\rm s}$ and $U_0=2.74\times10^{6}\,{\rm s}^{-1}$, and taking $\Delta_2=2\pi\times0.005\,\mathrm{MHz},\, \Delta_3=2\pi\times100\,\mathrm{MHz},\,
\Delta_4=2\pi\times0.18\,\mathrm{MHz},\, \Omega_{{c}}=2\pi\times6.5\,\mathrm{MHz},$  and
$\mathcal{N}_a=3\times10^{10}\,\mathrm{cm^{-3}}$, we obtain $V_{g1}\approx2.20\times10^{-5}c$ and
$V_{g2}\approx2.28\times10^{-5}c$,
which means the two polarization components of the probe field propagate with ultraslow  and nearly equal group velocities.  Because $\delta\approx 0$ and $g_{\delta}\ll 1$, the second term in Eq.~(\ref{eqn7}) can be safely neglected.

%%%%%%%%%%%%%%%%%%%%%%%%%%%%%%%%%%%%%%%%%%

\end{document}